\documentclass[11pt,a4paper]{article}
\pdfoutput=1

\usepackage{amsmath}
\allowdisplaybreaks
\usepackage[T1]{fontenc}

\usepackage{xpatch}
\usepackage{jheppub}
\usepackage{cleveref}
\usepackage{psfrag}
\usepackage{slashed}
\usepackage{cancel}
\usepackage{lscape}
\usepackage{caption}
\usepackage{array}
\usepackage{graphicx}
\usepackage{subcaption}
\usepackage{multirow}
\usepackage{tabularx}
\usepackage{makecell}
\usepackage[utf8]{inputenc}
\usepackage{amsmath}
\usepackage{amssymb}
\usepackage{color}
\usepackage{slashed}
\usepackage{comment}
\usepackage{mathtools}
\usepackage{siunitx}[=v2]
\usepackage{tikz}
\usepackage{tikz-feynman}
\tikzfeynmanset{compat=1.1.0}

\definecolor{mypink}{RGB}{219, 48, 122}
\definecolor{mygreen}{rgb}{0,0.7,0}
\definecolor{raspberry}{rgb}{0.53,0.15,0.34}

\def\la{\langle}
\def\ra{\rangle}
\def\spA#1#2{\la#1#2\ra}
\def\spB#1#2{[#1#2]}

\DeclareMathOperator{\tr}{\rm tr}
\def\trm{\tr_-}

\def\eps{\epsilon}

\newcommand{\dd}{\mathop{}\!\mathrm{d}}

\newcommand{\eqn}[1]{Eq.\,(\ref{#1})}

\newcommand{\fig}[1]{Figure\,\ref{#1}}

\newcommand{\sct}[1]{Section\,\ref{#1}}

\newcommand{\app}[1]{Appendix\,\ref{#1}}
\def\citere#1{\mbox{Ref.\,\cite{#1}}}

\usepackage{relsize}
\newcommand{\muR}{\mu_{\text{\relscale{0.77}R}}}
\newcommand{\Litwo}{\text{Li}_2}
\usepackage[normalem]{ulem}
\newcommand{\CYMloop}{C^{\phi{(0)}\times \text{YM}{(1)}}}
\newcommand{\Cphiloop}{C^{\phi{(1)}\times \text{YM}{(0)}}}
\newcommand{\CgenphiiYMj}{C^{\phi{(i)}\times \text{YM}{(j)}}}
\newcommand{\FtwoMS}{F'^{(2)}}
\newcommand{\FtwoccMS}{F'^{(2),cc}}

\def\hpl11{{\mathrm{HPL}}_{1,1}}

\newcolumntype{C}[1]{>{\hsize=#1\hsize\centering\arraybackslash}X}%

\newcolumntype{Z}{r<{\hspace{3mm}}}

\title{Two-loop all-plus helicity amplitudes for self-dual Higgs boson with gluons via unitarity cut constraints}

\author[a]{Simon Badger,}
\author[b]{Christian Biello,}
\author[a]{Colomba Brancaccio,}
\author[a]{Federico Ripani}

\affiliation[a]{
Dipartimento di Fisica and Arnold-Regge Center, Universit\`{a} di Torino,
and INFN, Sezione di Torino, Via P. Giuria 1, I-10125 Torino, Italy
}
\affiliation[b]{
Institute for Theoretical Physics, ETH Zurich, 8093 Zürich, Switzerland
}

\emailAdd{
simondavid.badger@unito.it, cbiello@phys.ethz.ch, colomba.brancaccio@unito.it, federico.ripani@unito.it
}

\abstract{
We present the two-loop amplitudes for a self-dual Higgs boson with up to four positive helicity gluons in the heavy top-quark limit. Because the tree
amplitudes in the all-plus sector vanish, we can construct simple representations of the polylogarithmic parts of the two-loop amplitudes using
four-dimensional unitarity cuts into rational one-loop and tree amplitudes. The remaining rational function ambiguity is extracted from a tensor integral reduction over finite fields.
The final expressions are presented using polylogarithms up to weight two and compact rational functions of spinor-helicity products.
}
\keywords{}
\preprint{}

\begin{document}
\maketitle
\flushbottom

\section{Introduction}

The self-dual Higgs model, in which the complex scalar field $\phi$ couples to gluons through the self-dual component of the gluonic field strength, exhibits remarkable simplifications in its on-shell tree-level amplitudes~\cite{Dixon:2004za}. The construction was used to find compact analytic expressions for the one-loop Higgs plus four
parton amplitudes~\cite{Badger:2006us,Berger:2006sh,Badger:2007si,Badger:2009hw,Badger:2009vh,Dixon:2009uk}. Indeed, a computation of partons plus the self-dual Higgs boson for all possible helicities is equivalent to computing both Higgs and pseudoscalar Higgs production with QCD jets.

An unusual feature of this model, unlike in standard Higgs amplitudes in the heavy-top limit, is that the tree-level amplitudes with all-plus gluon helicities or with a single minus-helicity gluon vanish. This mirrors the well-known pattern in Yang–Mills theory, where such amplitudes are protected by supersymmetric Ward identities~\cite{Grisaru:1976vm,Grisaru:1977px,Parke:1985pn}. A similar argument can be extended to the self-dual theory~\cite{Dixon:2004za}, although it strictly applies in the limit of a massless Higgs field.

Scalar Higgs plus four parton amplitudes at two loops in QCD are a current priority ingredient~\cite{Huss:2025nlt,Andersen:2024czj} for experimental measurements of Higgs boson
properties. Recently developed techniques, such as amplitude reduction over finite fields and the construction of special-function bases for finite (IR- and UV-subtracted) remainders, are now mature enough to be applied to these amplitudes. They also enable a detailed exploration of the all-plus sector in the self-dual Higgs model and its comparison to the corresponding sector in Yang–Mills theory, where complete expressions with up to seven external legs have been computed~\cite{Dalgleish:2024sey} and
all-multiplicity structure have been identified for the logarithmic and dilogarithmic terms~\cite{Dunbar:2020wdh}. In addition, bootstrapping
approaches in the all-plus sector of gauge theories have been recently investigated. For example, it has been shown that an identification between
form factors in self-dual gauge theory and correlators in a celestial chiral algebra can be used to bootstrap some two-loop all-plus
amplitudes~\cite{Costello:2023vyy,Dixon:2024mzh,Morales:2025alm}. Further study in this direction has found connections with the linear relations in Yang-Mills all-plus
and single-minus partial amplitudes~\cite{Dixon:2024tsb}.

Over the past decade, major advances have been made in computational methods for two-to-three scattering amplitudes, addressing both analytic (Feynman
integrals, special functions) and algebraic (kinematic coefficients) challenges. Key developments include optimized integration-by-parts (IBP)
reduction techniques~\cite{Gluza:2010ws,Ita:2015tya,Larsen:2015ped,Wu:2023upw,Guan:2024byi} and improved analytic understanding of multi-scale Feynman integrals through differential equations~\cite{Gehrmann:2015bfy,Papadopoulos:2015jft,Abreu:2018rcw,Chicherin:2018mue,Chicherin:2018old,Abreu:2018aqd,Abreu:2020jxa,Canko:2020ylt,Abreu:2021smk,Kardos:2022tpo,Abreu:2023rco}. These efforts enabled the
classification of relevant integrals into bases of “pentagon functions”, which efficiently represent infrared- and ultraviolet-subtracted finite
remainders~\cite{Gehrmann:2018yef,Chicherin:2020oor,Chicherin:2021dyp,Abreu:2023rco} (see also refs.~\cite{Badger:2023xtl,FebresCordero:2023pww,Gehrmann:2024tds}). Rational coefficients of these functions can now be extracted numerically using finite-field techniques~\cite{vonManteuffel:2014ixa,Peraro:2016wsq,Klappert:2019emp,Peraro:2019svx,Smirnov:2019qkx,Klappert:2020aqs,Klappert:2020nbg}, avoiding cumbersome intermediate
expressions. This framework has led to numerous new results for double-virtual QCD
corrections~\cite{Badger:2019djh,Badger:2021nhg,Badger:2021ega,Abreu:2021asb,Agarwal:2021vdh,Badger:2021imn,Abreu:2023bdp,Badger:2023mgf,Agarwal:2023suw,DeLaurentis:2023nss,DeLaurentis:2023izi,Badger:2022ncb,Badger:2024sqv,Badger:2024mir,DeLaurentis:2025dxw,Agarwal:2024jyq},
many of which have been successfully combined with real radiation contributions to produce high-precision predictions for differential cross sections~\cite{Chawdhry:2019bji,Kallweit:2020gcp,Chawdhry:2021hkp,Czakon:2021mjy,Badger:2021ohm,Chen:2022ktf,Alvarez:2023fhi,Badger:2023mgf,Hartanto:2022qhh,Hartanto:2022ypo,Buonocore:2022pqq,Catani:2022mfv,Buonocore:2023ljm,Mazzitelli:2024ura,Devoto:2024nhl,Biello:2024pgo,Buccioni:2025bkl}.

The study we present here is motivated by some observations of previously identified analytic structures. The fact that one-loop amplitudes for all-plus configurations are rational means that we may directly obtain a cut-constructible part of the two-loop amplitudes from unitarity~\cite{Bern:1994zx,Bern:1994cg}, or
generalised unitarity cuts~\cite{Bern:1997sc,Britto:2004nc}. By construction, the form of these cut-constructible pieces is one-loop-like and so will contain a maximum of weight-two polylogarithmic functions. In Yang-Mills theory, this structure is not manifest until the construction of the UV and IR subtracted finite remainders,
the loop integrals appearing in the amplitude are genuinely two-loop like and so contain up to weight four functions. Observing the weight drop in the function basis for five-point integrals with an off-shell leg is a useful test of the reduction technology applicable to phenomenologically relevant amplitudes. However, this is not our main motivation, as it is not immediately clear exactly how the cancellation occurs in the self-dual theory and in what way the choice of regularisation scheme plays a role. It may even be possible to determine an all-multiplicity form of the two-loop amplitude in this
sector, following a similar program to the one followed in Yang-Mills
theory~\cite{Dunbar:2016aux,Dunbar:2016gjb,Dunbar:2019fcq,Dunbar:2020wdh,Dalgleish:2020mof,Kosower:2022bfv,Dunbar:2023ayw,Dalgleish:2024sey}.

We recall that in Yang-Mills theory, the cut-constructible terms could be separated from the
rational function using an expansion in the spin dimension parameter $d_s=g^\mu_{\ \mu}$,
\begin{equation}
  A^{(L)} = \sum_{k=0}^L (d_s-2)^k A^{(L),(d_s-2)^k} \,,
  \label{eq:dsexpansion}
\end{equation}
where the $d_s$ dependence of the amplitude comes from the contraction of the metric tensor throughout each loop.
The expansion in $d_s-2$ is useful for pure Yang-Mills since we may see it as a sort of super-symmetric decomposition in which the first term
vanishes for all-plus (and single minus) configurations. At two loops, it was observed that $A^{(2),(d_s-2)}$ contained all cut-constructible terms
while $A^{(2),(d_s-2)^2}$ was purely rational and contained all the terms not accessible via unitarity cuts in four dimensions. The computation of
$A^{(2),(d_s-2)^2}$ is considerably simpler than $A^{(L)}$ since it only involves one-loop squared integral topologies and hence only one-loop integration
and reduction.

The fact that the one-loop amplitudes for the self-dual Higgs with positive helicity gluons are rational means that there must be a weight drop for the finite
remainders. However, the argument for the $d_s$ expansion no longer holds, and so it's interesting to see how the components are related to the
cut-constructible expression and whether there is a simple one-loop type computation that could be used to fix the rational ambiguity. If any such
pattern can be observed, there would be potential to extend computations to higher multiplicity.

Our paper is organised as follows: We first review the self-dual Higgs model and recall the known amplitudes that we will use to construct the two-loop
expression. In Section \ref{sec:cuts}, we outline how generalised unitarity cuts can be used to identify all logarithmic and dilogarithmic terms. In Section \ref{sec:finrem} we discuss the structure of the universal IR and UV poles and define the finite remainders for the direct diagram computation
in the `t Hooft-Veltman (tHV) scheme~\cite{THOOFT1972189} (in which $d_s=4-2\epsilon$) and for the four-dimensional cut-constructible part. This allows us to define a rational remainder function which
is determined by sampling the diagrams over finite fields as described in Section \ref{sec:rat}. In particular, we explore the connection between the
rational remainder function and the one-loop squared topologies included in $A^{(2),(d_s-2)^2}$. In \sct{sec:coll}, we perform checks of the expected factorisation in the double and triple collinear limits before drawing our conclusions.

\section{Review of the self-dual Higgs model \label{sec:rev}}

The Lagrangian for the self-dual Higgs field ($\phi$) coupling to gluons in the heavy quark limit is,  
\begin{equation}
  \mathcal{L} = C\left( \phi \, {\rm tr} G_{SD, \mu\nu}G_{SD}^{\mu\nu} + \phi^\dagger \, G_{ASD, \mu\nu}G_{ASD}^{\mu\nu}\right)\,.
  \label{eq:selfdualLag}
\end{equation}
Here, we introduce the self-dual and anti-self-dual component of the gluonic field strength $G^{\mu\nu}$ as follows:
\begin{equation}
  G^{\mu\nu}_{SD} = \frac{1}{2}\left(G^{\mu\nu} + \tilde{G}^{\mu\nu}\right)\,, \quad
  G^{\mu\nu}_{ASD} = \frac{1}{2}\left(G^{\mu\nu} - \tilde{G}^{\mu\nu}\right)\,,
  \label{eq:sdfieldstrength}
\end{equation}
with
\begin{equation}
	\tilde{G}^{\mu\nu} = \frac{i}{2} \varepsilon^{\mu\nu\rho\sigma}G_{\rho\sigma}\,,
  \label{eq:fieldstrength}
\end{equation}
using $\varepsilon^{\mu\nu\rho\sigma}$ to represent the totally antisymmetric Levi-Civita symbol in four dimensions.

We will consider colour-ordered amplitudes of the self-dual Higgs field, $\phi$, with gluons in the leading-colour approximation,
\begin{equation}
  -i \mathcal{A}^{(L)}(\phi;1,\dots,n) = \mathcal{N}^L g_s^{(n-2)} C \sum_{\xi \in \mathbb{S}_n/\mathbb{Z}_n} {\rm tr}(t^{a_{\xi(1)}}t^{a_{\xi(2)}} \dots t^{a_{\xi(n)}}) A^{(L)}(\phi;\xi(1),\dots,\xi(n))\,.
  \label{eq:colourdecomp}
\end{equation}
We have chosen standard normalisations of the loop integrals and coupling constants with $\mathcal{N} = m_\eps N_c \alpha_s/(4\pi)$, $m_\eps = (4\pi)^\eps e^{-\eps\gamma_E}$, $\alpha_s = g_s^2/(4\pi)$ ($\gamma_E$ is the Euler-Mascheroni constant) and $t^{a_{i}}$ is the generator of $\text{SU}(N_c)$ in the fundamental representation. The sum over colour factors, $\mathbb{S}_n/\mathbb{Z}_n$, runs over all non-cyclic permutations of the gluons. The (leading order) Wilson coefficient of the effective dimension five operator is $C = \alpha_s/(6\pi v)$ where $v$ is the Higgs vacuum
expectation value. The kinematics is defined in~\app{app:kinematics}, while the non-standard Feynman rules are described in~\app{app:feynrules}.

We will use the previously known results for tree and one-loop amplitudes inside the (four-dimensional) cuts. At tree-level~\cite{Dixon:2004za},
\begin{equation}
  \begin{aligned}
  A^{(0)}(\phi;1^+,2^+,\dots,n^+) &= 0\,, \\
  A^{(0)}(\phi^\dagger;1^-,2^+,3^+) &= \frac{{\spB32}^3}{\spB31\spB12}\,, \\
  A^{(0)}(\phi;1^-,2^+,\dots,n^+) &= 0 \ \ (n>3)\,, \\
  A^{(0)}(\phi^\dagger;1^+,2^+,\dots,n^+) &= \frac{s_\phi^2}{\spA12\spA23\dots\spA{n}1}\,,
  \label{eq:trees}
  \end{aligned}
\end{equation}
where $s_\phi = p_\phi^2$ is the invariant mass of the Higgs boson (not necessarily on-shell). At one-loop level~\cite{Berger:2006sh},
\begin{equation}
  \begin{aligned}
  A^{(1)}(\phi;1^+,2^+,\dots,n^+) &= -2 A^{(0)}( \phi^\dagger;1^+,2^+,\dots,n^+)\,,\\
  A^{(1)}(\phi;1^-,2^+) &= 0\,, \\
  A^{(1)}(\phi;1^-,2^+,3^+) &= \frac{{\spB32}^3}{\spB31\spB12}\left( -2+\frac{s_{12}s_{13}}{3 s_{23}^2} \right)\,.
  \label{eq:1loop}
  \end{aligned}
\end{equation}
Note that above, we have taken into account the contribution from gluon loops (and ghost loops, depending on the gauge choice), i.e. we drop closed fermion-loop terms. When considering the universal IR pole structure at two loops, higher-order terms in the $\eps$ expansion at one loop are required.

In this work we do not consider fermion fields. In principle the same method
considered here could also be used to compute the closed fermionic loop
contributions to the processes presented in this paper. In fact the structure
is somewhat simpler than the gluonic loops since the one-loop fermionic loop
contribution to the all-plus configuration are zero hence the cut-constructible
contributions at two-loops are also zero. The two-loop contributions to the
finite remainders are expected rational and could be extracted via finite field
based reduction methods. An additional complication is that
the effective Lagrangian also contains a light quark axial current in the
presence of fermions. This term can contribute to the two-loop
corrections of the all-plus helicity amplitudes with effective one-loop
diagrams, not included in our setup. Since there is no phenomenological
motivation for these contributions, and the fact that cut-constructible terms
are zero, we choose to leave this computation for the future.


\section{The cut-constructible contributions \label{sec:cuts}}

In this section, we present the cut-constructible part of the amplitude, obtained by evaluating the unitarity cuts in four dimensions. Based on the optical theorem and the Cutkosky analysis of scattering amplitude discontinuities~\cite{Cutkosky:1960sp,Eden:1966dnq}, modern unitarity methods~\cite{Bern:1994zx,Bern:2005hs,Bern:2005ji,Bern:2005cq} have been developed as powerful computational techniques to reconstruct amplitudes from knowledge of a suitable integral basis. For a recent pedagogical description of the cut-construction method, we refer the reader to~\citere{Badger:2023eqz}. Since the tree-level amplitude in the self-dual Higgs theory vanishes for configurations with all-plus gluons, leading to the absence of triple cuts, the corresponding two-loop amplitude can be expressed in terms of one-loop scalar integrals, known as the cut-constructible part, together with a purely rational component free of branch cuts. The coefficients of the scalar integrals can be systematically determined by imposing on-shell conditions on the two-loop propagators and isolating the relevant terms using spinor-helicity algebra. The cut-constructible contribution can therefore be reconstructed by analysing the double cuts formed by the product of tree-level (\eqn{eq:trees}) and rational one-loop (\eqn{eq:1loop}) subamplitudes discussed in the previous section. We present the explicit calculation for the $\phi + 2$ gluon amplitude, with additional details for higher multiplicities provided in~\app{app:ccdetails}.

By applying two on-shell conditions on the loop momentum, the resulting cut integral can be interpreted as a linear combination of scalar integrals with two cut propagators. For $\phi+2$ gluons, we have the sum of two double-cut contributions in the only non-trivial channel:
\begin{align}
	&\text{Disc}_{s_\phi}\left[A^{(2)}(\phi;1^+,2^+)\right] = \sum_{\sigma\in \mathbb{Z}_2}  \int  \dd \Phi_2 \sum_{h_i=\pm} \left[ \Cphiloop_{s_\phi}(\phi;\sigma(1)^+,\sigma(2)^+,\ell_1^{h_1},\ell_2^{h_2})  \right. \nonumber \\
	&\hspace{0.5cm}\left.+\CYMloop_{s_\phi}(\phi;\sigma(1)^+,\sigma(2)^+,\ell_1^{h_1},\ell_2^{h_2}) \right] \label{eq:dssphi}\,,
\end{align}
where the integrands are defined as follows,
\begin{align}
	&\CgenphiiYMj_{s_\phi}(\phi;\sigma(1)^+,\sigma(2)^+,\ell_1^{h_1},\ell_2^{h_2}) \equiv \\
	&\hspace{0.5cm} A^{(i)}(\phi;(-\ell_1)^{-h_1},\ell_2^{h_2}) A^{(j)}(\ell_1^{h_1},(-\ell_2)^{-h_2},\sigma(1)^+,\sigma(2)^+)\,, \label{def:CCintegrand}
\end{align}
while the usual on-shell phase-space for double cuts is
\begin{align}
	\dd \Phi_2 =\frac{\dd^4 \ell_1}{(2\pi)^4}\frac{\dd^4 \ell_2}{(2\pi)^4} (2\pi)^4 \delta^{(4)}(\ell_1-\ell_2-p_\phi) (2\pi)\delta^{(+)}(\ell_1^2) (2\pi)\delta^{(+)}(\ell_2^2)\,.
\end{align}
Owing to the colourless nature of the self-dual Higgs field, the partial amplitude is obtained by considering all possible insertions of the scalar among the gluons, or equivalently, by summing over all the cyclic permutations of the $n$ external gluons generated by group elements $\sigma \in\mathbb{Z}_n$.

\begin{figure}[h]
  \centering
  \includegraphics[width=0.4\textwidth, page=1]{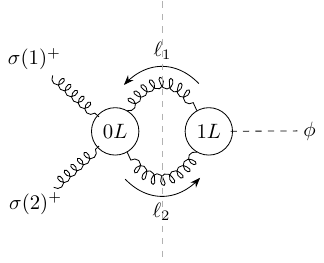}
  \hspace{0.1cm}
  \includegraphics[width=0.4\textwidth, page=2]{fig/Phi+2g-tikzext.pdf}
  \caption{The two double cuts in the $s_\phi$ channel of the $\phi+2g^+$ amplitude.}
  \label{fig:dcutsphi2g}
\end{figure}

The two integrand contributions of \eqn{eq:dssphi} are represented in~\fig{fig:dcutsphi2g}. Due to the helicity structure of tree-level Yang-Mills and $\phi$ amplitudes (see \eqn{eq:trees}), we only need the MHV amplitudes for the tree-level inputs in the integrand terms, together with all-plus one-loop amplitudes. The calculation is straightforward:
\begin{align}
	&\Cphiloop_{s_\phi}(\phi;1^+,2^+,\ell_1^-,\ell_2^+) = -2A^{(0)}(\phi^\dagger;1,2) \frac{\langle 12\ell_1 \ell_2 1 ]}{\langle 1 \ell_2 1] \langle 2 \ell_1 2]}\,, \label{dcutC1}\\
	&\CYMloop_{s_\phi}(\phi;1^+,2^+,\ell_1^+,\ell_2^-) = -\frac{1}{3} A^{(0)}(\phi^\dagger;1,2)\,. \label{eq:dcutC2}
\end{align}

The integrand in \eqn{eq:dcutC2} does not depend anymore on the loop momentum, therefore it can be interpreted as the result of a double cut of a scalar bubble. The first coefficient requires the removal of spurious terms that integrate to zero in order to show a correspondence with cut scalar integrals. The $\gamma_5$ contribution of $\trm(1,2,\ell_1,\ell_2)=\langle 12 \ell_1 \ell_2 1]$ is spurious, therefore by expanding the trace one can show that
\begin{align}
	\Cphiloop_{s_\phi}(\phi;1^+,2^+,\ell_1^-,\ell_2^+) = s_{\phi} A^{(0)}(\phi;1,2) \left[ \frac{1}{2 p_1\cdot \ell_2} - \frac{1}{2 p_2\cdot \ell_1}\right] +\dots\,,
\end{align}
where the dotted part is vanishing after integration. By summing over the two possible permutations $\sigma$, and using the symmetry of the on-shell phase space $\dd\Phi_2$, we obtain
\begin{align}
	&\text{Disc}_{s_\phi}\left[A^{(2)}(\phi;1^+,2^+)\right] = \nonumber \\
	& \hspace{0.5cm} 4 s_\phi A^{(0)}(\phi^\dagger;1^+,2^+) \text{Disc}_{s_\phi}\left[ I_3^{1m}(s_\phi) \right] - \frac{2}{3} A^{(0)}(\phi^\dagger;1^+,2^+)  \text{Disc}_{s_\phi}\left[ I_2^{}(s_\phi) \right] \,.
\end{align} 
Here, we introduce the scalar integrals with massless propagators as defined in Ref.~\cite{Ellis:2007qk}. The only difference is the overall normalisation, based here on the $\mathcal{N}$ factor of the partial amplitude decomposition in \eqn{eq:colourdecomp}. With the subscript, we denote the number of external legs, while the superscript is used to specify the number of massive external states. We therefore define the cut-constructable part of the $\phi+2g$ amplitude as follows:
\begin{align}
	A^{(2),cc}(\phi;1^+,2^+)&=\left[-2A^{(0)}(\phi^\dagger;1^+,2^+)\right] \left[-2 s_\phi I_3^{1m}(s_\phi)  + \frac{1}{3} I_2^{}(s_\phi)  \right] \\
	&=\left[-2A^{(0)}(\phi^\dagger;1^+,2^+)\right]\left[-\frac{2}{\epsilon^2}+\frac{1}{\epsilon}\left(\frac{1}{3}-2\log\frac{\muR^2}{-s_\phi-i\delta}\right) \right. \nonumber\\
	&\hspace{0.3cm}\left.+\frac{\pi^2}{6}+\frac{2}{3} +\frac{1}{3}\log\frac{\muR^2}{-s_\phi-i\delta}-\log^2\frac{\muR^2}{-s_\phi-i\delta} +\mathcal{O}(\epsilon) \right] \,.
\end{align}
In the previous equation, we have introduced the renormalisation scale $\muR$ and the small positive parameter $\delta=0^+$ as a prescription for the analytic continuation.

\begin{figure}[t]
  \centering
  \includegraphics[width=0.4\textwidth, page=1]{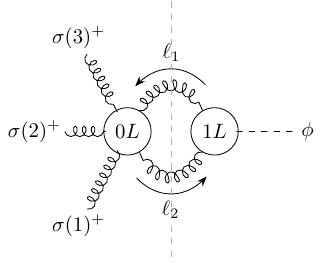}
  \hspace{0.1cm}
  \includegraphics[width=0.4\textwidth, page=2]{fig/Phi+3g-tikzext.pdf}\\
    \hspace{0.38cm}\includegraphics[width=0.4\textwidth, page=3]{fig/Phi+3g-tikzext.pdf}
  \hspace{0.1cm}
  \includegraphics[width=0.4\textwidth, page=4]{fig/Phi+3g-tikzext.pdf}\\
  \caption{The double cuts of the $\phi+3g^+$ amplitude.}
  \label{fig:dcut3-sphi}
\end{figure}

In the higher-multiplicity case, we have to compute the double cuts for several channels, requiring additional algebraic simplifications. Nevertheless, the cut-constructible approach remains conceptually straightforward, with the possibility of bootstrapping parts of the amplitude by expressing them in terms of scalar one-loop integrals. In~\fig{fig:dcut3-sphi}, for example, we present the double cut contributions we have computed to determine the discontinuity of the $\phi+3g$ amplitude. In~\app{app:ccdetails}, we explicitly report the simplified integrands of the double cuts for all channels in the $\phi+3g$ and $\phi+4g$ amplitudes. Here, we present the resulting cut-constructible expressions, starting with the $\phi+3g$ amplitude:
\begin{align}
	&A^{(2),cc}(\phi;1^+,2^+,3^+)= \left[-2A^{(0)}(\phi^\dagger;1^+,2^+,3^+)\right] \left[-\frac{1}{\epsilon^2}\sum_{i=1}^3 \left( \frac{\muR^2}{-s_{ii+1}-i\delta} \right)^\epsilon +\frac{1}{3\epsilon} \right]+ \nonumber \\
        &\hspace{0.4cm}  \left[-2A^{(0)}(\phi^\dagger;1^+,2^+,3^+)\right] \left[\frac{3\pi^2}{4}+\frac{2}{3}+\frac{1}{3}\log\frac{\muR^2}{-s_\phi-i\delta}+\frac{1}{2} \log^2\left(\frac{-s_{12}-i\delta}{-s_{23}-i\delta}\right)  \right.\nonumber\\
        &\hspace{0.4cm} \left.+\frac{1}{2} \log^2\left(\frac{-s_{23}-i\delta}{-s_{12}-i\delta}\right) +\frac{1}{2} \log^2\left(\frac{-s_{13}-i\delta}{-s_{12}-i\delta}\right)+ \text{Li}_2 \left(1-\frac{-s_\phi-i\delta}{-s_{12}-i\delta}\right)\right. \nonumber\\
        &\hspace{0.4cm} \left. + \text{Li}_2 \left(1-\frac{-s_\phi-i\delta}{-s_{23}-i\delta}\right) + \text{Li}_2 \left(1-\frac{-s_\phi-i\delta}{-s_{13}-i\delta}\right)\right] \,.\label{eq:A2ccphi3g}
\end{align}

The cut-constructible part of the $\phi+4g$ amplitude is more intricate. For clarity, we decompose it into two distinct contributions: one arising from the set of double cuts involving the rational one-loop $\phi$ amplitudes and the tree-level pure-gluonic amplitudes, denoted as $A^{(2),cc}_{\phi{(1)}\times \text{YM}{(0)}}$, and the complementary configuration where the roles are reversed, $A^{(2),cc}_{\phi{(0)}\times \text{YM}{(1)}}$. The full cut-constructible contribution is thus given by:
\begin{align}
	& A^{(2),cc}(\phi;1^+,2^+,3^+,4^+)= A^{(2),cc}_{\phi{(1)}\times \text{YM}{(0)}}+ A^{(2),cc}_{\phi{(0)}\times \text{YM}{(1)}}\,.\label{eq:A2ccphi4g}
\end{align}
The first sector yields the following expression in terms of polylogarithms:
\begin{align}
	&A^{(2),cc}_{\phi{(1)}\times \text{YM}{(0)}}=\left[-2A^{(0)}(\phi^\dagger;1^+,2^+,3^+,4^+)\right] \times \Bigg\{ -\frac{1}{\epsilon^2}\sum_{i=1}^4 \left( \frac{\muR^2}{-s_{ii+1}-i\delta} \right)^\epsilon  \nonumber \\
	&\hspace{0.2cm} + \sum_{\sigma\in \mathbb{Z}_4} \left[ 2 \text{Li}_2\left(1-\frac{-s_\phi-i\delta}{-s_{\sigma(1)\phi}-i \delta}\right) - \Litwo\left( 1-\frac{(-s_\phi-i\delta) (-s_{\sigma(1)\sigma(2)}-i\delta)}{(-s_{\sigma(3)\phi}-i\delta)(-s_{\sigma(4)\phi}-i\delta)}\right)\right. \nonumber\\
	&\left.-\,\eta\left(\frac{-s_\phi-i\delta}{-s_{\sigma(3)\phi}-i\delta},\frac{-s_{\sigma(1)\sigma(2)}-i\delta}{-s_{\sigma(4)\phi}-i\delta}\right) \log\left( 1-\frac{(-s_\phi-i\delta) (-s_{\sigma(1)\sigma(2)}-i\delta)}{(-s_{\sigma(3)\phi}-i\delta)(-s_{\sigma(4)\phi}-i\delta)}\right)\right. \nonumber\\
	&\left. +\frac{1}{2}\ln^2\left( \frac{-s_{\sigma(1)\phi}-i\delta}{-s_{\sigma(2) \phi}-i\delta} \right)+ \frac{1}{2}\ln^2\left( \frac{-s_{\sigma(1)\sigma(2)}-i\delta}{-s_{\sigma(2) \sigma(3)}-i\delta} \right) - \frac{1}{2}\ln^2\left( \frac{-s_{\sigma(1)\phi}-i\delta}{-s_{\sigma(2) \sigma(3)}-i\delta} \right) \right. \nonumber\\
	&\left. - \frac{1}{2}\ln^2\left( \frac{-s_{\sigma(1)\phi}-i\delta}{-s_{\sigma(3) \sigma(4)}-i\delta} \right)\right]+\frac{2 \pi ^2}{3} \Bigg\} \,, \label{eq:A2ccfirstsector}
\end{align}   
where
\begin{align}
	\eta(x,y)=\log(xy)-\log(x)-\log(y) 
\end{align}
is needed for capturing the proper analytical continuation of some contributions from two-mass scalar boxes. We stress that some straightforward simplifications between polylogarithmic contributions of different scalar integrals occur once we sum over all the possible cyclic permutations of the scalar field. The second set of double cuts provides the following cut-constructable contribution:
\begin{align}
   &A^{(2),cc}_{\phi{(0)}\times \text{YM}{(1)}}=-
	\sum_{\sigma\in\mathbb{Z}_4}\frac{1}{3}\frac{[\sigma(2)\sigma(3)]^2}{\langle \sigma(1)\sigma(4)\rangle^2}\left[
\text{Li}_2\left(1-\frac{-s_\phi-i\delta}{-s_{\sigma(1)\phi}-i\delta}\right) \right. \nonumber\\&\left. +\text{Li}_2\left(1-\frac{-s_\phi-i\delta}{-s_{\sigma(4)\phi}-i\delta}\right)
+\text{Li}_2\left(1-\frac{-s_{\sigma(2)\sigma(3)}-i\delta}{-s_{\sigma(1)\phi}-i\delta}\right) 
+\text{Li}_2\left(1-\frac{-s_{\sigma(2)\sigma(3)}-i\delta}{-s_{\sigma(4)\phi}-i\delta}\right) \right. \nonumber\\
&\left. +\frac{1}{2}\ln^2\left(\frac{-s_{\sigma(1)\phi}-i\delta}{-s_{\sigma(4)\phi}-i\delta}\right)-\text{Li}_2\left(1-\frac{(-s_\phi-i\delta)(- s_{\sigma(2)\sigma(3)}-i\delta)}{(-s_{\sigma(1)\phi}-i\delta)(-s_{\sigma(4)\phi}-i\delta)}\right)\right. \nonumber\\
&\left.-\eta\left(\frac{-s_\phi-i\delta}{-s_{\sigma(1)\phi}-i\delta},\frac{- s_{\sigma(2)\sigma(3)}-i\delta}{-s_{\sigma(4)\phi}-i\delta}\right)\log\left(1-\frac{(-s_\phi-i\delta)(- s_{\sigma(2)\sigma(3)}-i\delta)}{(-s_{\sigma(1)\phi}-i\delta)(-s_{\sigma(4)\phi}-i\delta)}\right)
\right] \nonumber
\\&+\frac{1}{3}\left[-2A^{(0)}(\phi^\dagger;1^+,2^+,3^+,4^+)\right] \left[\frac{1}{\epsilon}+2+\ln\left(\frac{\muR^2}{-s_\phi-i\delta}\right)\right]\,. \label{eq:A2ccsecondsector}
\end{align}

The box contributions in the previous cut-constructible amplitudes have been confirmed by applying quadruple generalised cuts~\cite{Bern:1997sc,Britto:2004nc,Bern:2007dw}.


\section{Finite remainders and universal pole structure \label{sec:finrem}}

To check the cut-constructible expression presented in the previous section matches a direct computation, we perform a reduction of the relevant
Feynman diagrams to special functions using finite field evaluations. In addition, we will be able to determine the non-cut-constructible
contributions, missing from the four-dimensional unitarity analysis. We may first check the poles of the cut-constructible expression against the expected form. Logarithms of soft-collinear origin are simple, but there are some subtle points concerning the $1/\eps$ poles.

To subtract the UV singularities, we have to renormalise the strong coupling constant $\alpha_s$ and the self-dual operator
$\operatorname{tr}(G_{SD,\mu\nu}G^{\mu\nu}_{SD})$. The self-dual operator is a linear combination of $\operatorname{tr}(G_{\mu\nu}G^{\mu\nu})$ and
$\operatorname{tr}(G_{\mu\nu}\tilde{G}^{\mu\nu})$, which have a different renormalisation at two-loop level. This difference is manifest in the
presence of a counter term proportional to the tree-level $\phi^\dagger$ amplitude in renormalised two-loop $\phi$ amplitudes. The clearest way to see
it is to decompose $\phi$ in its scalar, $H=\phi+\phi^\dagger$ and pseudo-scalar, $H_{\rm pseudo} = -i(\phi-\phi^\dagger)$, components, namely, in terms of renormalised amplitudes,
\begin{equation}
  \mathcal{A}_{\text{ren}}(\phi) = \frac{1}{2}\mathcal{A}_{\text{ren}}(H)+\frac{i}{2}\mathcal{A}_{\text{ren}}(H_{\rm pseudo}),
\label{Aren}
\end{equation} 
where
\begin{equation}
  \mathcal{A}_{\text{ren}}(H) = Z_{GG}\mathcal{A}(H), \;\;\;\;\;\;\;\;\; \mathcal{A}_{\text{ren}}(H_{\rm pseudo}) = Z_{G\tilde{G}}\mathcal{A}(H_{\rm pseudo}),
\end{equation}
and renormalisation of $\alpha_s$ is understood. Mixing of $\operatorname{tr}(G_{\mu\nu}\tilde{G}^{\mu\nu})$ with axial anomaly operator enters in the
$N_f$ contribution and so does not affect the partial amplitudes considered in this paper. The renormalisation constant $Z_{GG}$ has been known for a
long time~\cite{Kluberg-Stern,Spiridonov} in $\overline{\text{MS}}$-scheme up to two-loop level, while $\overline{\text{MS}}$ $Z_{G\tilde{G}}$ is
given at the same perturbative order in Ref.~\cite{Larin_1993} in tHV scheme~\cite{THOOFT1972189}. 
For completeness, we give here the expressions used to build the counter term in the leading-colour limit: 
\begin{equation}
\label{ZGG}
  \begin{aligned}
     Z_{GG} & = 1 + \sum_{L=1}^{\infty} \mathcal{N}^L \delta Z_{GG}^{(L)}\,, \quad \delta Z_{GG}^{(1)} = 
     -\frac{11}{3\eps}\,, \quad \delta Z_{GG}^{(2)} = \frac{121}{9\eps^2} - \frac{34}{3\eps}\,,\\
     Z_{G\tilde{G}} & = 1 + \sum_{L=1}^{\infty} \mathcal{N}^L \delta Z_{G\tilde{G}}^{(L)}\,, \quad \delta Z_{G\tilde{G}}^{(1)}
      = -\frac{11}{3\eps}\,, \quad \delta Z_{G\tilde{G}}^{(2)} = \frac{121}{9\eps^2} - \frac{17}{3\eps}\,,\\
     Z_{\alpha_s} & = 1 + \sum_{L=1}^{\infty} \mathcal{N}^L \delta Z_{\alpha_s}^{(L)}\,,
   \quad \delta Z_{\alpha_s}^{(1)} = -\frac{11}{3\eps}\,,
   \end{aligned}
\end{equation}
where $\mathcal{N}$ has been defined in Section \ref{sec:rev}.
After expanding Eq.~\eqref{Aren} in the strong coupling constant, one- and two-loop renormalised $\phi+n$ gluon amplitudes, in the all-plus helicity configuration, read
\begin{equation}
\label{A12ren}
  \begin{aligned}
  A^{(1)}_{\text{ren}}(\phi) & = A^{(1)}(\phi) \,, \\
  A^{(2)}_{\text{ren}}(\phi) & = A^{(2)}(\phi) + \delta Z_{\alpha_s}^{(1)}\left(\frac{n}{2}+1\right)A^{(1)}(\phi) 
  + \frac{1}{2}\left(\delta Z_{GG}^{(2)} - \delta Z_{G\tilde{G}}^{(2)}\right)A^{(0)}(\phi^\dagger) \,,
  \end{aligned}
\end{equation} 
where we used the fact that the tree-level $\phi$ amplitude vanishes, together with the relation
\begin{equation}
  \delta Z_{GG}^{(1)} = \delta Z_{G\tilde{G}}^{(1)} = \delta Z_{\alpha_s}^{(1)} \,.
\end{equation}

The IR structure is captured by the usual Catani's formula \cite{Catani_1998}, where the contribution proportional to the tree-level amplitude is null
for the all-plus helicity configuration. Finally, subtracting the IR divergences from~\eqn{A12ren} and using the explicit form for the renormalisation
constants, we define the finite remainder, in the tHV scheme, as follows,
\begin{equation}
  \begin{aligned}
  F^{(2)}(\phi;1^+,2^+,\dots,n^+) =
  &A^{(2),[4-2\eps]}(\phi;1^+,2^+,\dots,n^+) \\
  - &I^{(1)} A^{(1),[4-2\eps]}(\phi;1^+,2^+,\dots,n^+)\\
  - &J^{(2)} A^{(0)}(\phi^\dagger;1^+,2^+,\dots,n^+) \,,
  \label{eq:finremHV}
\end{aligned}
\end{equation}
where
\begin{equation}
  I^{(1)} = -\frac{1}{\eps^2}\sum_{i=1}^n \left( \frac{\muR^2}{-s_{i i+1}-i\delta} \right)^\eps + \frac{11}{3\eps} \,,
  \label{eq:I1HV}
\end{equation}
and
\begin{equation}
  J^{(2)} = \frac{17}{6\eps} - \frac{151}{12} - \log\left(\frac{\muR^2}{-s_{\phi}-i\delta} \right)\,.
  \label{eq:J2HV}
\end{equation}
Note that we explicitly include the dimensional regularisation label on the amplitudes to highlight that higher-order terms in $\eps$ are required in $A^{(1),[4-2\eps]}$. The finite contribution to $J^{(2)}$ is introduced in order to nullify the rational remainder~\eqref{eq:ratrem} for the two-gluon amplitude.

The cut-constructible expression we derived does not match up exactly with this form. We define,
\begin{equation}
  F^{(2),cc}(\phi;1^+,2^+,\dots,n^+) = A^{(2),cc}(\phi;1^+,2^+,\dots,n^+) - I^{(1),cc} A^{(1)}(\phi;1^+,2^+,\dots,n^+) \,,
  \label{eq:finremCC}
\end{equation}
where
\begin{equation}
  I^{(1),cc} = -\frac{1}{\eps^2}\sum_{i=1}^n \left( \frac{\muR^2}{-s_{i i+1}-i\delta} \right)^\eps + \frac{1}{3\eps}\,.
  \label{eq:I1CC}
\end{equation}
Initially, the difference in the $1/\eps$ pole seems alarming, but analysing $D$-dimensional cuts for low multiplicity cases shows that one can miss
terms like $\eps/\eps^2$ with four-dimensional cuts. We also note that in the finite remainder of the cut-constructible expression higher-order terms in $\eps$ for the one-loop amplitude are not required.

To complete our computation of the amplitudes with up to four gluons, we will determine the difference,
\begin{equation}
  R^{(2)}(\phi;1^+,2^+,\dots,n^+) = F^{(2)}(\phi;1^+,2^+,\dots,n^+)-F^{(2),cc}(\phi;1^+,2^+,\dots,n^+)\,. 
  \label{eq:ratrem}
\end{equation}

The analytic expressions for the finite remainder of the cut-constructible contribution, defined in \eqn{eq:finremCC}, and of the rational
contribution, defined in \eqn{eq:ratrem}, are provided in \texttt{Wolfram Mathematica} language in the ancillary files accompanying this article~\cite{zenodo}. Furthermore, the expressions for $R^{(2)}(\phi;1^+,2^+,\dots,n^+)$, with $2 \leq n \leq 4$, are given in the next section, see \eqn{eq:results}, together with a more detailed discussion of their derivation.

\section{The rational remainder \label{sec:rat}}

The simplest way to proceed is perhaps to compute $A^{(2),[4-2\eps]}$ directly using the `t Hooft-Veltman prescription for $\gamma_5$. While this is
at the limit of current technology, it is feasible, particularly bearing in mind that the final expressions are much simpler than the general helicity
cases. We choose to generate the integrand using Feynman diagrams, which puts additional strain on this approach since the Feynman rules contain a lot
of extra terms, and cancellations are even more prevalent than in Yang-Mills theory.

In any case, such an approach would not scale to higher multiplicity in the near future, and so we choose to explore the structure of the expansion in
$d_s-2$ discussed in the introduction. For Yang-Mills amplitudes,

\begin{equation}
  R^{(2)}(1^+,2^+,\dots,n^+) = 4 R^{(2),(d_s-2)^2}(1^+,2^+,\dots,n^+)
  \label{eq:YMrem}
\end{equation}
where $R^{(2),(d_s-2)^2}$ contains only one-loop squared topologies and can be studied using generalised unitarity cuts in
$D$-dimensions~\cite{Kosower:2022bfv}. This is a dramatic simplification, and the number of diagrams required to compute the remainder function is reduced.

\subsection{Tensor integral reduction setup \label{ssec:red}}

In the following, we describe the workflow employed for the analytic reconstruction of the two-loop $\phi + 2g$ and $\phi + 3g$ amplitudes, as well as
for the $(d_s - 2)^2$ component of the two-loop $\phi + 4g$ amplitude ($A^{(2),(d_s-2)^2}(\phi,1^+,2^+,3^+,4^+)$) and the tHV two-loop $\phi + 4g$ amplitude evaluated over finite fields. Our workflow builds upon the codebase developed and applied in several previous amplitude computations~\cite{Badger:2021nhg,Badger:2021ega,Badger:2022ncb,Badger:2021imn,Badger:2023mgf,Badger:2024sqv,Badger:2024awe,Badger:2024dxo}.

\begin{figure}[h]
  \centering
  \includegraphics[width=\textwidth, page=1]{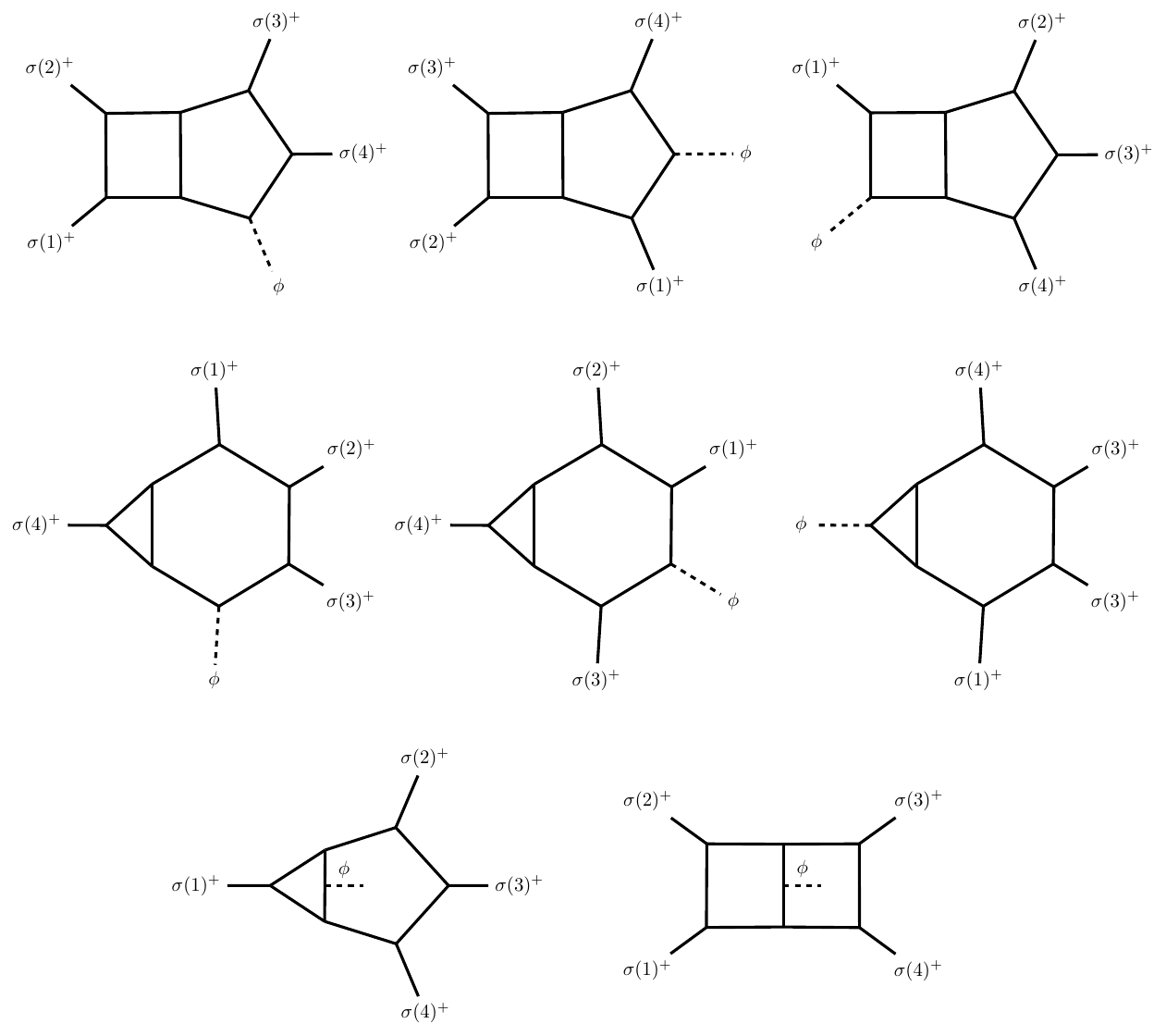}
  \caption{The maximal cut topologies contributing to $\phi+4g$ in the leading colour limit.
Black lines denote massless particles and the dashed line the massive scalar particle $\phi$.}
  \label{fig:maxcuttopo}
\end{figure}

The first step consists in generating the Feynman diagrams representing the targeted process using the software \texttt{QGRAF}~\cite{Nogueira:1991ex} with the Feynman rules defined in~\app{app:feynrules}. Subsequently, we perform the colour decomposition, retaining only the leading-colour contribution in the $1/N_c$ expansion (see \eqn{eq:colourdecomp}). In the leading-colour approximation, the number of diagrams contributing to the $\phi + 4g$ amplitude amounts to 7348, organised in 1007 topologies. We then carry out the Lorentz contractions and map all loop-integral topologies onto a choice of maximal-cut topologies. To this aim, we map bubble insertions onto hexagon-triangle topologies. For the leading-colour $\phi + 4g$ amplitude, we find 52 maximal-cut topologies which correspond, after identifying different permutations of massless particles, to 8 maximal-cut topologies, illustrated in~\fig{fig:maxcuttopo}. The steps described so far are implemented using the package \texttt{SPINNEY}~\cite{Cullen:2010jv} and a combination of in-house \texttt{Mathematica} and \texttt{FORM}~\cite{Ruijl:2017dtg} scripts.

We build a table of rules to map the numerators to polynomials in the inverse propagators and irreducible scalar products (ISPs) of the corresponding topologies, where the coefficients depend only on external kinematics, specifically on momentum-twistor variables \cite{Hodges:2009hk, Badger:2016uuq}. The chosen momentum-twistor parametrisation is defined in~\app{app:kinematics}. At this stage, we sample momentum-twistor variables and dimensional regulator $\epsilon$ over finite fields $\mathbb{Z}_p$, the set of integers modulo a prime $p$, replacing symbolic calculations with finite-field arithmetic~\cite{vonManteuffel:2014ixa,Peraro:2016wsq} employing the framework \texttt{FiniteFlow}~\cite{Peraro:2019svx}. 

The two-loop amplitude, expressed in terms of scalar Feynman integrals, is subsequently decomposed into the master integrals (MIs) basis introduced in Ref.~\cite{Abreu:2023rco} by performing an integration-by-parts (IBP) reduction. For the four-gluon case, we have 65573 tensor integrals with a maximum rank of 6 and at most one dotted propagator. The IBP identities are generated in an optimised way exploiting the syzygy method \cite{Gluza_2011,Chen_2016,B_hm_2018,Larsen_2018,bendle2020,Larsen_2016,zhang2018} by using the \texttt{NeatIBPs} package~\cite{Wu:2023upw}. In particular, we set up a system of IBP identities for each maximal-cut topology, shown in~\fig{fig:maxcuttopo} for the four-gluon case, and for only one permutation of the external legs. Reduction of permuted topologies is performed by solving multiple times the same system for the different permutations of the Mandelstam variables $s_{ij}$, see \eqn{eq:sijtomt}, for their expression in terms of momentum-twistor variables. After the reduction, symmetry relations between MIs of all the topologies and permutations are applied, leading to a drop in the number of MIs in the $\phi + 4g$ amplitude from 3422 to 844. Both the amplitude and the MIs are expanded around $\epsilon = 0$, and the MIs components of the expansion are mapped onto the special-function basis introduced in Ref.~\cite{Abreu:2023rco}. Since the all-plus $\phi +$gluons amplitude vanishes at tree level, the two-loop amplitude presents a simpler functional structure at two loops compared to other helicity configurations. However, the transcendental weight drop becomes apparent only after mapping the MIs onto the special-function basis and subtracting IR and UV singularities. Up to this stage, the algebraic complexity of the amplitude remains comparable to that of other helicity configurations.

In order to reduce the maximal polynomial degree of the expression evaluated over finite fields that needs to be reconstructed, we apply univariate
partial fraction decomposition~\cite{Badger:2021nhg}. We stress that the denominators in the rational coefficients of the special functions are known
by matching them to the letters corresponding to the MIs. In particular, we perform the functional reconstruction only of the $(d_s-2)^2$ component of
the two-loop $\phi+4g$ amplitude, $A^{(2),(d_s-2)^2}(\phi,1^+,2^+,3^+,4^+)$. As illustrated in Table~\ref{tab:coefficients}, we observe that the
maximal polynomial degree of the special function coefficients of $A^{(2),(d_s-2)^2}(\phi,1^+,2^+,3^+,4^+)$ are comparable to the ones of the tHV
two-loop amplitude $A^{(2)}(\phi,1^+,2^+,3^+,4^+)$,\footnote{Recall the tHV scheme is defined by setting $d_s=4-2\epsilon$.} but the evaluation time for a
single numerical evaluation over finite fields is a factor of $100$ faster compared to tHV amplitude. 

\begin{table}[h!]
\centering
\begin{tabular}{|c|c|c|c|}
\hline
\textbf{Amplitude/max degree} & \textbf{MIs coefficients} & \textbf{SF coefficients} & \textbf{UPFD} \\ \hline
$A^{(2),(d_s-2)^2}(\phi,1^+,2^+,3^+,4^+)$ & $96/85$ & $46/33$ & $25/0$ \\ \hline
$A^{(2)}(\phi,1^+,2^+,3^+,4^+)$ & $223/211$ & $47/34$ & $25/0$ \\ \hline
\end{tabular}
\caption{Data about the functional reconstruction of $A^{(2),(d_s-2)^2}(\phi,1^+,2^+,3^+,4^+)$ and $A^{(2)}(\phi,1^+,2^+,3^+,4^+)$. From the second to the fourth column, we show the maximal numerator/denominator polynomial degree at the various stages of the computation. In particular, we show the maximal polynomial degree of the coefficients of the amplitude written in terms of MIs, in terms of special functions and the degree after performing univariate partial fraction decomposition.}
\label{tab:coefficients}
\end{table}

\subsection{Results for up to four gluons}

The extraction of the rational remainder from the $(d_s-2)^2$ diagrams was possible, but did not follow the simple structure observed for Yang-Mills
theory. The finite remainder of $A^{(2),(d_s-2)^2}(\phi,1^+,2^+,3^+,4^+)$ is not purely rational and also contained logarithms and di-logarithms. These
functions can be easily written in terms of the one-loop integral functions observed in the double cuts. In contrast to the coefficients of the
cut-constructible terms identified in~\sct{sec:cuts}, this amplitude contains spurious singularities of the form
$\tfrac{\log(s_{i\phi}/s_\phi)}{s_{i\phi}-s_\phi}^2$ where $i$ is one of the gluon momenta. These spurious singularities at $s_{i\phi}\to s_\phi$ cancel with some of
the rational (weight zero) terms. The procedure to complete the spurious poles in the weight one functions with weight zero functions is well known
and has been used in many one-loop examples (see for example~\cite{Bern:1993mq,Bern:1994fz}) and also for the single-minus Yang-Mills two-loop amplitudes~\cite{Badger:2018enw}. It is therefore straightforward to extract
a (much simpler) rational term without spurious poles that matches the pole structure of the rational remainder in the tHV scheme. This feature can be
identified using a univariate slice modulo a prime field, which is quick to obtain. In the case of the $\phi+3g$ amplitude, the rational remainder is
proportional to the rational term for the $(d_s-2)^2$ amplitude once spurious poles were removed. For the $\phi+4g$ amplitude, the partial fractioning
of spurious poles was not unique, and $(d_s-2)^2$ didn't match up exactly with the (slice mod prime) of the tHV amplitude. Nevertheless, a simple
structure of terms could be identified, which formed a suitable ansatz for the tHV rational remainder, such that it could then be determined through
$\mathbb{Q}$-linear relations (all computations modulo a prime field). A compact set of rational remainders was obtained using spinor-helicity
techniques,
\begin{equation}
  \begin{aligned}
  R^{(2)}(\phi;1^+,2^+) &= 0\,, \\
  R^{(2)}(\phi;1^+,2^+,3^+) &= -\frac{2}{3}\frac{1}{\spA12\spA23\spA31} \sum_{\sigma\in\mathbb{Z}_3} s_{\sigma(1)\sigma(2)} s_{\sigma(2)\sigma(3)}\,, \\
  R^{(2)}(\phi;1^+,2^+,3^+.4^+) &= -\frac{1}{3}\frac{1}{\spA12\spA23\spA34\spA41} \sum_{\sigma\in\mathbb{Z}_4} \Bigg(s_{\sigma(1)\sigma(2)} s_{\sigma(2)\sigma(3)}\\&
  + \frac{1}{4} (s_{\sigma(1)\phi}+s_{\sigma(3)\phi})(s_{\sigma(2)\phi}+s_{\sigma(4)\phi})
  + \frac{1}{12} s_{\sigma(1)\sigma(3)} s_{\sigma(2)\sigma(4)}\\&
  - \frac{1}{6} s_{\sigma(1)\sigma(2)} s_{\sigma(3)\sigma(4)}
  - \frac{1}{6} \frac{\trm(\sigma(2)\sigma(3)\sigma(4)\sigma(1))^2}{s_{\sigma(1)\sigma(2)}s_{\sigma(3)\sigma(4)}} \\&
   - \frac{(s_{\sigma(1)\sigma(2)}+s_\phi)\trm(\sigma(2)\sigma(3)\sigma(4)\phi)}{s_{\sigma(3)\sigma(4)}}\\&
  + \frac{s_{\sigma(1)\sigma(3)}(s_{\sigma(1)\sigma(2)}+s_\phi)\trm(\sigma(2)\sigma(3)\sigma(4)\phi)}{s_{\sigma(2)\phi}s_{\sigma(3)\sigma(4)}}\\&
  - \frac{\trm(\sigma(2)\sigma(3)\sigma(4)\phi)^2}{s_{\sigma(2)\phi}s_{\sigma(3)\sigma(4)}}  - \frac{\trm(\sigma(1)\sigma(2)\phi\,\sigma(4))^2}{s_{\sigma(2)\phi}s_{\sigma(1)\sigma(4)}}\\&
  + \frac{s_{\sigma(1)\sigma(3)}(s_{\sigma(2)\sigma(3)}+s_\phi)\trm(\sigma(1)\sigma(2)\phi\,\sigma(4))}{s_{\sigma(2)\phi}s_{\sigma(1)\sigma(4)}}
  \Bigg)\,.
  \label{eq:results}
  \end{aligned}
\end{equation}

\section{Collinear limit checks \label{sec:coll}}

We have verified that our expressions satisfy the expected factorisation in both double~\cite{Bern:1999ry} and triple collinear limits~\cite{Badger:2015cxa,Czakon:2022fqi}.

To ensure explicit factorisation of the finite remainder, we define
\begin{equation}
  \FtwoMS(\phi;1^+,2^+,\dots,n^+) =  R^{(2)}(\phi;1^+,2^+,\dots,n^+) + \FtwoccMS(\phi;1^+,2^+,\dots,n^+)\,,
  \label{eq:F2Coll}
\end{equation}
with
\begin{equation}
  \FtwoccMS(\phi;1^+,2^+,\dots,n^+) = F^{(2),cc}(\phi;1^+,2^+,\dots,n^+) + I^{(1),cc}_{fin} A^{(1)}(\phi;1^+,2^+,\dots,n^+),
  \label{eq:F2ccColl}
\end{equation}
where $I^{(1),cc}_{fin}$ denotes the $\epsilon^0$ term of $I^{(1),cc}$~\eqref{eq:I1CC}. More precisely, it takes the following form:
\begin{align}
	I^{(1),cc}_{fin} = -\frac{1}{2}  \sum_{i=1}^4 \log^2\left(\frac{\muR^2}{-s_{ii+1}-i\delta}\right)\,.
\end{align}
Indeed, we add back to the finite remainder of the cut-constructible terms the finite contribution of the Catani operator, such that we can verify the factorisation for the minimally-subtracted amplitude. In this way, we can directly use the $\mathcal{O}(\epsilon^0)$ contributions of the universal un-renormalised splitting amplitudes to construct the expected factorisation of $\FtwoMS$.

Since the tree-level collinear splitting amplitude ${\rm Sp}^{(0)}(1^+,2^+;-P_{12}^+)$ and one-loop finite reminder $F^{(1)}(\phi;1^-,2^+)$ are vanishing, the $\phi+3g$ amplitude
has a simple factorisation in the double collinear limit $1||2$,
\begin{equation}
  \begin{aligned}
    \FtwoMS(\phi;1^+,2^+,3^+) \overset{1||2}{\rightarrow}
    &\FtwoMS(\phi;P_{12}^+,3^+) {\rm Sp}^{(0)}(1^+,2^+;-P_{12}^-) \\
  + &F^{(1)}(\phi;P_{12}^+,3^+) {\rm Sp}^{(1)}(1^+,2^+;-P_{12}^-)\,.
  \label{eq:phi3gcoll12}
  \end{aligned}
\end{equation}
A pictorial representation of this factorisation formula is shown in Fig.~\ref{fig:collinear1}.
\begin{figure}[h]
  \centering
  \includegraphics[width=1.0\textwidth, page=1]{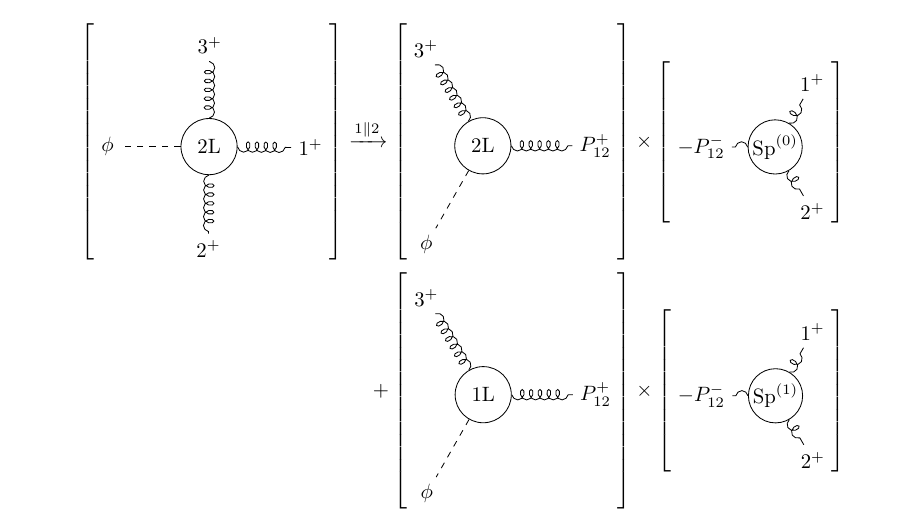}
  \caption{Pictorial representation of the factorisation~\eqref{eq:phi3gcoll12}.}
  \label{fig:collinear1}
\end{figure}

Moreover, the splitting functions introduced in~\eqn{eq:phi3gcoll12} read~\cite{Bern:1999ry}
\begin{equation}
  \begin{aligned}
    {\rm Sp}^{(0)}(1^+,2^+;-P_{12}^-) = & \,\frac{1}{\spA{1}{2}}\frac{1}{\spA{z_1}{}\spA{z_2}{}} \,, \\
    {\rm Sp}^{(1)}(1^+,2^+;-P_{12}^-) = & \left( \frac{1}{3}\frac{1}{z_1 z_2} 
	 - \big( \log\!\left(z_1\right)
	 + \log\!\left(z_2\right) \big)
   \log\!\left(\frac{\muR^2}{- s_{12}-i\,\delta} \right)
	-\frac{1}{2}\,\log^2\!\left(\frac{\muR^2}{- s_{12} -i\,\delta} \right)
	   	\right. \\
   & \left.
		-\frac{1}{2}\,\log^2\!\left(
  \frac{\spA{z_1}{}\spB{z_1}{}}
       {\spA{z_2}{}\spB{z_2}{}}
	\right)
	-\frac{\pi^2}{6}
 \right) {\rm Sp}^{(0)}(1^+,2^+;-P_{12}^-) \,,
  \label{eq:spl++-}
  \end{aligned}
\end{equation}
where
\begin{align}
  \spA{z_i}{} = \frac{\spA{i}{n}}{\spA{P}{n}}\,, \qquad
  \spB{z_i}{} = \frac{\spB{i}{n}}{\spB{P}{n}}\,, \qquad
  z_i = \spA{z_i}{}\spB{z_i}{}.
  \label{eq:zdef}
\end{align}
In \eqn{eq:zdef}, we introduced the momentum \( P \) defined as \( P = p_1 + p_2 \), 
and the arbitrary light-like reference vector \( n \) satisfying \( n^2 = 0 \). The variables \( z_1 \) and \( z_2 \) represent the momentum fractions of the combined system \( 1+2 \) carried by \( p_1 \) and \( p_2 \), respectively, with \( p_i = z_i P \). We stress that $\spA{z_1}{}\spB{z_1}{}+\spA{z_2}{}\spB{z_2}{}=1$, which reflects the momentum conservation in the collinear limit. The spinor-helicity notation adopted for the splitting amplitudes is inspired by~\cite{Badger:2015cxa}, to which we refer for a more detailed discussion. Since we defined the rational component as vanishing in the $\phi+2$ gluon case, the factorisation of $R^{(2)}(\phi;1^+,2^+,3^+)$ matches only the rational part of the second term of the RHS in~\eqn{eq:phi3gcoll12}, while the first term involves only the cut-constructible pieces. We have analytically verified the factorisation in \eqn{eq:phi3gcoll12} by employing simple dilogarithm identities that reduce in the collinear limit the cut-constructible part to logarithmic terms only, together with spinor-helicity algebra.

The double collinear limit of the $\phi+4g$ amplitude has a slightly more complicated structure since $F^{(1)}(\phi;1^-,2^+,3^+)$ is non-zero,
\begin{equation}
  \begin{aligned}
  \FtwoMS(\phi;1^+,2^+,3^+,4^+) \overset{1||2}{\rightarrow}
    &\FtwoMS(\phi;P_{12}^+,3^+,4^+) {\rm Sp}^{(0)}(1^+,2^+;-P_{12}^-) \\
  + &F^{(1)}(\phi;P_{12}^+,3^+,4^+) {\rm Sp}^{(1)}(1^+,2^+;-P_{12}^-) \\
  + &F^{(1)}(\phi;P_{12}^-,3^+,4^+) {\rm Sp}^{(1)}(1^+,2^+;-P_{12}^+)\,.
  \label{eq:phi4gcoll12}
  \end{aligned}
\end{equation}

\begin{figure}[h]
  \centering
  \includegraphics[width=1.0\textwidth, page=2]{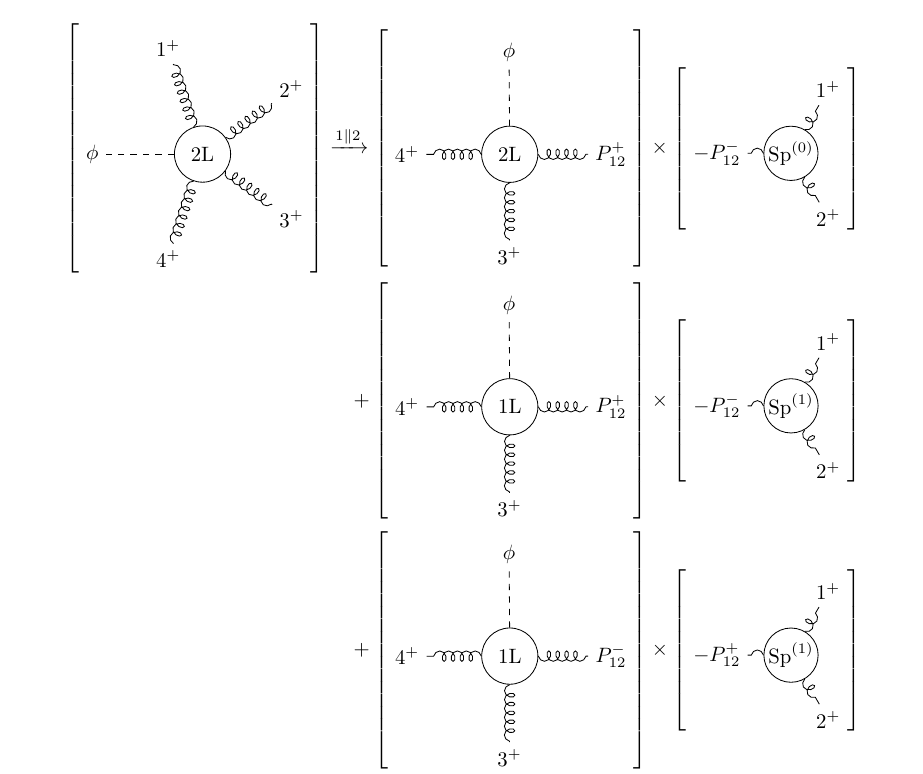}
  \caption{Pictorial representation of the factorisation~\eqref{eq:phi4gcoll12}.}
  \label{fig:collinear1}
\end{figure}

Since ${\rm Sp}^{(0)}(1^+,2^+;-P_{12}^+)$ is zero, ${\rm Sp}^{(1)}(1^+,2^+;-P_{12}^+)$ is rational. Therefore, the last term of~\eqn{eq:phi4gcoll12} is only sensitive to the factorisation behaviour of $R^{(2)}(\phi;1^+,2^+,3^+,4^+)$. In particular, ${\rm Sp}^{(1)}(1^+,2^+;-P_{12}^+)$ reads as 
\begin{equation}
  \begin{aligned}
	{\rm Sp}^{(1)}(1^+,2^+;-P_{12}^+) =  \frac{1}{3} \frac{\spA{z_1}{}\spA{z_2}{}\spB{1}{2}}{\spA{1}{2}^2}.
  \label{eq:spl+++}
  \end{aligned}
\end{equation}
We have verified the factorisation in \eqn{eq:phi4gcoll12}. We have observed that the factorisation of the cut-constructible part follows a similar pattern to what is observed in the collinear factorisation of one-loop Yang-Mills and $\phi^\dagger$ amplitudes~\cite{Bern:1994zx,Badger:2007si}, where box and triangle integrals with off-shell legs are mapped into simpler scalar integrals in the collinear limit in order to match the cut-constructible component at lower multiplicity. We stress that, although we have written the factorisation in a compact and unified form, the cut-constructible and rational components each factorise independently into their corresponding lower-multiplicity contributions.

Finally, we describe the check of the triple collinear limit, which is only possible for the $\phi+4g$ amplitude,
\begin{equation}
  \begin{aligned}
  \FtwoMS(\phi;1^+,2^+,3^+,4^+) \overset{1||2||3}{\rightarrow}
    &\FtwoMS(\phi;P_{123}^+,4^+) {\rm Sp}^{(0)}(1^+,2^+,3^+;-P_{123}^-) \\
  + &F^{(1)}(\phi;P_{123}^+,4^+) {\rm Sp}^{(1)}(1^+,2^+,3^+;-P_{123}^-) \,.
  \label{eq:phi4gcoll123}
  \end{aligned}
\end{equation}

\begin{figure}[h]
  \centering
  \includegraphics[width=1.0\textwidth, page=3]{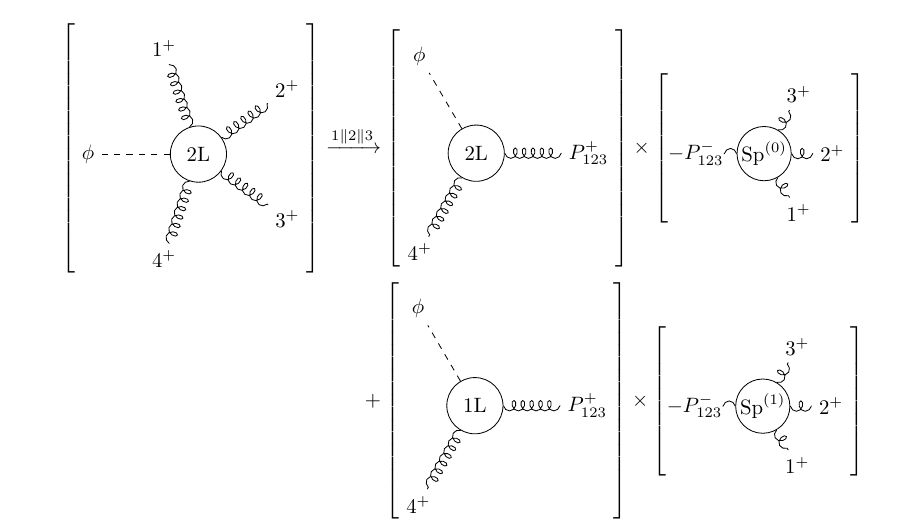}
  \caption{Pictorial representation of the triple-collinear limit of the $\phi+4$ gluon amplitude~\eqref{eq:phi4gcoll123}.}
  \label{fig:collinear1}
\end{figure}

It is sensitive to the tree-level and one-loop triple collinear splitting
amplitudes~\cite{Badger:2015cxa}:
\begin{equation}
  \begin{aligned}
  {\rm Sp}^{(0)}(1^+,2^+,3^+;-P_{123}^-)  = &\, \frac{1}{\spA{z_1}{} \spA{z_3}{}} 
  				\frac{1}{\spA{1}{2}\spA{2}{3}} \,, \\
  {\rm Sp}^{(1)}(1^+,2^+,3^+;-P_{123}^-) = &\, {\rm Sp}^{(0)}(1^+,2^+,3^+;-P_{123}^-) \left( V_{cc} + V_R\right)  \,,
  \end{aligned}
\end{equation}
with
\allowdisplaybreaks
\begin{align}
 V_{cc} = &
    - \text{Li}_2\left(1-\frac{-s_{23}-i\delta}{-(1-z_1) s_{123}-i\delta}\right)-
   \text{Li}_2\left(1-\frac{-s_{12}-i\delta}{-(1-z_3) s_{123}-i\delta}\right)\nonumber\\
   &-\frac{1}{2} \log
   ^2\left(\frac{\muR^2}{-s_{12}-i \delta} \right)-\frac{1}{2} \log ^2\left(\frac{\muR^2}{-s_{23}-i \delta} \right)- \log \left(\frac{-s_{12}-i\delta}{-s_{123}-i\delta}\right) \log
   \left(\frac{-s_{23}-i\delta}{-s_{123}-i\delta}\right)\nonumber\\
   &+ \log \left(\frac{-s_{12}-i\delta}{-s_{123}-i\delta}\right) \log
   \left(\frac{1-z_3}{z_1}\right)+ \log
   \left(\frac{-s_{23}-i\delta}{-s_{123}-i\delta}\right) \log
   \left(\frac{1-z_1}{z_3}\right)\nonumber\\
   &- \log (z_1) \log
   \left(\frac{\muR^2}{-s_{123}-i \delta }\right)- \log (z_3) \log
   \left(\frac{\muR^2}{-s_{123}-i \delta }\right)+
   \text{Li}_2\left(-\frac{z_2}{z_1}\right)\nonumber\\
   &+
   \text{Li}_2\left(-\frac{z_1}{1-z_1}\right)+
   \text{Li}_2\left(-\frac{z_2}{z_3}\right)+
   \text{Li}_2\left(-\frac{z_3}{1-z_3}\right)+\frac{\pi ^2}{6} \,,\\
 V_R = &\, \frac{1}{3} \left( z_1 z_2 + z_1 z_3 + z_2 z_3
	+ \frac{z_1 z_2 (1 - z_2^2) z_3}
       {(1 - z_1)(1 - z_2)(1 - z_3)}  
	   	\right. \nonumber\\
   & \left.
 	- z_1 z_3 \spB{2}{P} \left(
    \frac{\spB{z_1}{}}{(1 - z_3)\spB{1}{2}}
    - \frac{\spB{z_3}{}}{(1 - z_1)\spB{2}{3}}
  \right) 
	+ \frac{\spB{z_1}{}\spB{z_3}{}\spB{2}{P}^2 \, s_{13}}
       {\spB{1}{2}\spB{2}{3}\, s_{123}} \right) \,.
  \label{eq:VccVR}
\end{align}
Here, $s_{ij \hdots k} = (p_i + p_j + \hdots + p_k)^2$, while the momentum-fraction variables $z_i$ are defined according to~\eqn{eq:zdef}. We observe that the rational component factorises directly into the rational part $V_R$ of the one-loop splitting function, as the cut-constructible part has been defined such that the $\phi+2$ gluon amplitude contains no two-loop rational remainder. The cut-constructible component, in turn, factorises into a term proportional to the finite two-loop two-gluon remainder and the $V_{cc}$ part of the one-loop splitting function. We have verified the factorisation in the triple collinear limit in \eqn{eq:phi4gcoll123}.

Since a direct computation was possible, the collinear limits only serve as a cross-check of the result. Alternatively, one could also envisage a complementary approach, in which an ansatz for the rational remainder is constructed to satisfy all (multi-)collinear limits. However, we find that the collinear limit of the $\phi + 4$ gluon amplitude is insensitive to the component proportional to $\trm(1234)$, which appears in both the cut-constructible and rational parts at high multiplicity.

\section{Conclusions}

In this article, we have computed the two-loop QCD corrections to a self-dual Higgs with up to four positive helicity gluons for the first time.
Since the tree-level amplitudes vanish in this case, it was possible to obtain compact analytic representations for all cut-constructible contributions
from two-particle cuts in four dimensions. The four-dimensional cuts missed a rational function, which was computed directly through the reduction of
the two-loop Feynman diagrams over finite fields. We also checked that the final results are consistent with double and triple collinear limits. 

These amplitudes display some similarities to the all-plus amplitudes in Yang-Mills theory, which are now known to very high multiplicity. In the
Yang-Mills case, it is observed that all rational terms missed with four-dimensional cuts came from one-loop squared topologies proportional to
$(d_s-2)^2$. In this case, the connection between the $(d_s-2)^2$ component and the extraction of the rational remainder was less straightforward owing to the
appearance of spurious poles and logarithms. Nevertheless, we were able to use the $(d_s-2)^2$ subset of graphs to build an ansatz for the rational
remainder in the tHV scheme, which dramatically reduced the computational cost. Exploiting this structure may allow the computation of higher multiplicity
amplitudes, perhaps also leveraging the collinear limits to constrain the anstatz.

Possible future developments include employing generalised $D$-dimensional unitarity cuts, following a similar strategy to that of~\citere{Kosower:2022bfv} for Yang-Mills amplitudes. Moreover, given its elegant analytical structure, this theory may provide a convenient framework for developing methods to reconstruct higher-multiplicity amplitudes through recursion relations, in line with the study on the all-plus five-gluon amplitude in~\citere{Dunbar:2016aux}.

The reduction technology applied in this case is the same that can be applied to the computation of the (scalar) Higgs plus four parton amplitudes at
two loops. This computation shows that those, phenomenologically relevant, amplitudes are easily within reach although the polynomial degree and
special function basis will be substantially more complicated and require significiant computing resources. Optimisation of the reconstruction
algorithm would likely be of great benefit in that case as demonstrated recently for the case of vector boson plus four partons in the leading colour
limit~\cite{DeLaurentis:2025dxw}.

\section*{Acknowledgements}
We thank Heribertus Bayu Hartanto and Simone Zoia for useful discussions and comments on the manuscript. S. Badger, C. Brancaccio and F. Ripani acknowledge funding from the Italian Ministry of Universities and Research (MUR) through FARE grant R207777C4R and through grant PRIN 2022BCXSW9. The work of C. Biello was supported in part by the Swiss National Science Foundation, grant 10001706.

\appendix

\section{Kinematics} \label{app:kinematics}

We construct the helicity amplitudes for the process
\[
0 \to g(p_1) + \hdots + g(p_{n}) + \phi(p_{\phi}),
\]
where \( g \) denotes a gluon, \( \phi \) the self-dual Higgs, and we consider amplitudes with \( 2 \leq n \leq 4 \).  
We work within the tHV scheme, with \( d = 4 - 2\epsilon \) space-time dimensions, where the Levi-Civita pseudotensor \( \varepsilon_{\mu\nu\rho\sigma} \) and the external momenta \( p_i \) are treated as four-dimensional. The external momenta satisfy the conservation law
\[
\sum_{i=1}^{n} p_i +p_\phi= 0,
\]
together with the on-shell conditions
\[
p_i^2 = 0 \quad (i = 1, \ldots, n), \qquad p_{\phi}^2 = s_{\phi}.
\]

In particular, the five-particle kinematics for the case \( n = 4 \) can be described in terms of six independent Mandelstam invariants,
\[
\vec{s} = (s_{12}, s_{23}, s_{34}, s_{4\phi}, s_{1\phi}, s_{\phi}),
\]
where \( s_{ij} = (p_i + p_j)^2 \).  
Additionally, we define the pseudoscalar invariant
\[
\mathrm{tr}_5 = 4\,i\, \varepsilon_{\mu\nu\rho\sigma}\, p_1^{\mu} p_2^{\nu} p_3^{\rho} p_4^{\sigma}
            = [12]\langle23\rangle[34]\langle41\rangle - [23]\langle34\rangle[41]\langle12\rangle \,.
\]

A parametrization based on the Mandelstam variables $\vec{s}\,$ necessarily involves $\mathrm{tr}_5 \equiv \sqrt{\Delta_5}$, with $\sqrt{\Delta_5}$ being an irreducible polynomial in $\vec{s}$. In order to rationalize both $\sqrt{\Delta_5}$ and the spinor brackets $\langle ij \rangle$ and $[ij]$, we make use of momentum-twistor variables~\cite{Hodges:2009hk, Badger:2016uuq}. In this representation, the Mandelstam invariants can be written explicitly as
\begin{equation}
  \begin{aligned}
s_{12} &= x_1 \,, \\
s_{23} &= x_1 x_4 \,, \\
s_{34} &= -\frac{x_1 \left( x_2 x_3 - x_4 - x_3 x_4 - x_2 x_3 x_4 + x_2 x_3 x_4 x_5 \right)}{x_2}\,, \\
s_{4\phi} &= x_1 x_6\,, \\
s_{1\phi} &= x_1 x_3 \left( x_2 - x_4 x_5 \right)\,, \\
s_{\phi} &= \frac{x_1 x_3 \left( x_2 - x_4 \right) \left( -x_4 + x_4 x_5 + x_6 \right)}{x_4} \,.
  \label{eq:sijtomt}
  \end{aligned}
\end{equation}

\section{Feynman Rules}  \label{app:feynrules}

In this appendix, we list the Feynman rules for the self-sual Higgs model since they are non-standard. These are given with the convention of all
particles outgoing,
\begin{equation}
  \begin{aligned}
  V(1_g,2_g,3_\phi) =& -i \, C \, \delta^{a_1 a_2} \left(
  g^{\mu_1\mu_2} p_1.p_2 - p_1^{\mu_2} p_2^{\mu_1}
  - \frac{1}{4}{\rm tr}(\gamma_5 \gamma^{\mu_1} \slashed{p}_1 \gamma^{\mu_2} \slashed{p}_2) \right) \,, \\
  V(1_g,2_g,3_g,4_\phi) =& C \, g_s f^{a_1 a_2 a_3} \big(
    g^{\mu_1 \mu_2} (p_1-p_2)^{\mu_3}
  + g^{\mu_2 \mu_3} (p_2-p_3)^{\mu_1}\\&
  + g^{\mu_3 \mu_1} (p_3-p_1)^{\mu_2}
  + \frac{1}{4}{\rm tr}(\gamma_5 \gamma^{\mu_1} \gamma^{\mu_2} \gamma^{\mu_3} \slashed{p}_4) \big) \,,\\
  V(1_g,2_g,3_g,4_g,5_\phi) =& \frac{i}{2} \, C \, g_s^2 \,
    T^{a_1 a_2 a_3 a_4} \left( g^{\mu_1\mu_2} g^{\mu_3\mu_4} + g^{\mu_2\mu_3} g^{\mu_1\mu_4} - 2 g^{\mu_1\mu_3} g^{\mu_2\mu_4}\right)\\&
  + T^{a_1 a_3 a_4 a_2} \left( g^{\mu_1\mu_3} g^{\mu_2\mu_4} + g^{\mu_3\mu_4} g^{\mu_1\mu_2} - 2 g^{\mu_1\mu_4} g^{\mu_2\mu_3}\right)\\&
  + T^{a_1 a_4 a_2 a_3} \left( g^{\mu_1\mu_4} g^{\mu_2\mu_3} + g^{\mu_2\mu_4} g^{\mu_1\mu_3} - 2 g^{\mu_1\mu_2} g^{\mu_3\mu_4}\right) \,,
  \end{aligned}
\end{equation}
where $T^{a_1 a_2 a_3 a_4} = {\rm tr}(t^{a_1}t^{a_2}t^{a_3}t^{a_4}) + {\rm tr}(t^{a_1}t^{a_4}t^{a_3}t^{a_2})$, and we use the conventions $p_i$ for the momenta, $a_i$ for adjoint $\text{SU}(N_c)$ indices, $\mu_i$ for Lorentz indices.


\section{Cut-constructible contributions at higher multiplicity} \label{app:ccdetails}
In this appendix, we report the integrand of the double cuts for $\phi+3g$ and $\phi+4g$ amplitudes, simplified in terms of scalar integrals with cut propagators. 

\subsection{Self-dual Higgs plus three gluon}
There are two classes of channels contributing to the discontinuity of the $\phi+3g$ amplitude. As in the $\phi+2g$ case, we can consider both the $s_\phi$ channel and the $s_{\phi i}$ channels. Each channel gives rise to two distinct integrands, corresponding to the product of a tree-level Yang–Mills amplitude with a one-loop $\phi$ amplitude, or vice versa. All four integrands are diagrammatically represented by~\fig{fig:dcut3-sphi}.

Using the same notation as in~\eqn{def:CCintegrand}, we have the following integrands after applying Schouten identities and removing spurious terms:
\begin{align}
	&\Cphiloop(\phi;1^+,2^+,3^+,\ell_1^{-},\ell_2^+)=\left[-2A^{(0)}(\phi;1^+,2^+,3^+)\right] \left[ \frac{s_{1\phi}-s_{\phi}}{2} \left.I_3^{2m}(s_{1\phi},s_{\phi})\right|_{s_{\phi}\text{-cut}} \right.\nonumber \\
	&\hspace{0.5cm} \left.  +\frac{s_{3\phi}-s_{\phi}}{2} \left.I_3^{2m}(s_{3\phi},s_{\phi})\right|_{s_{\phi}\text{-cut}}  -\frac{s_{1\phi}s_{3\phi}}{2} \left.I_4^{1m}(s_{1\phi},s_{3\phi};s_{\phi})\right|_{s_{\phi}\text{-cut}}  \right]\,,\\
	&\Cphiloop_{s_{1\phi}}(\phi;1^+,2^+,3^+,\ell_1^{-},\ell_2^+)=\left[-2A^{(0)}(\phi;1^+,2^+,3^+)\right] \left[ (s_{1\phi-s_\phi}) \left.I_3^{2m}(s_{1\phi},s_\phi) \right|_{s_{1\phi}\text{-cut}}  \right. \nonumber\\
	&\hspace{0.5cm} \left. -\frac{s_{1\phi}s_{3\phi}}{2} \left.I_4^{1m}(s_{1\phi},s_{3\phi};s_\phi)\right|_{s_{1\phi}\text{-cut}} -\frac{s_{1\phi}s_{2\phi}}{2} \left.I_4^{1m}(s_{1\phi},s_{2\phi};s_\phi)\right|_{s_{1\phi}\text{-cut}} \right]\,, \\
	&\CYMloop(\phi;1^+,2^+,3^+,\ell_1^{+},\ell_2^-)=0\,, \label{eq:C3}\\
	&\CYMloop(\phi;1^+,2^+,3^+,\ell_1^{+},\ell_2^-)=\frac{1}{3}\frac{s_\phi}{\langle 12\rangle \langle 23 \rangle \langle 31 \rangle }\left[\frac{s_{12}s_{13}}{s_{3\phi}-s_{\phi}}+\frac{s_{13}s_{\phi}}{s_{1\phi}-s_\phi}\right]\,. \label{eq:C4}
\end{align}
Here, the coefficients are expressed in terms of the integrands of scalar box and triangle functions, with the cut propagators indicated by the corresponding subscripts. For the triangle integrals, $I_3^{2m}(s_a, s_b)$, we denote the two invariants associated with the off-shell legs. For the box integrals, $I_4^{1m}(s, t; m^2)$, the arguments in parentheses correspond to the usual Mandelstam invariants and the off-shellness of the external legs. The vanishing behavior of~\eqn{eq:C3} can be shown, for example, by an integrand reduction of three-point tensor integrals with double-cut propagators. By summing over all possible permutations, the coefficient in~\eqn{eq:C4} simplifies to reproduce the same bubble term as in the lower-multiplicity case.

From the double cut analysis, we define the cut-constructible part of the self-dual Higgs plus three gluon amplitude as follows:
\begin{align}
	&A^{(2),cc}(\phi;1^+,2^+,3^+)= \left[-2 A^{(0)}( \phi^\dagger;1^+,2^+,3^+)\right]	\Big[\,\frac{1}{3} I_2(s_\phi)+ \nonumber\\
	&\hspace{0.5cm} + \sum_{\sigma\in \mathbb{S}_3} (s_{\phi\sigma(i)}-s_\phi) I_{3}^{2m}(s_{\phi\sigma(i)},s_\phi)- \sum_{\sigma\in \mathbb{S}_3} \frac{s_{\phi\sigma(i)}s_{\phi\sigma(i+2)}}{2} I_4^{1m}(s_{\phi\sigma(i)},s_{\phi\sigma(i+2)};s_\phi) \Big]\,.
\end{align}
The modulo-3 identification is implicit in the arguments of the permutation element $\sigma\in \mathbb{S}_3$. By expressing the scalar integrals in terms of (poly-)logarithms, the previous equation simplifies to~\eqn{eq:A2ccphi3g}.

\subsection{Self-dual Higgs plus four gluon}
\begin{figure}[h!]
  \centering
  \includegraphics[width=0.45\textwidth, page=1]{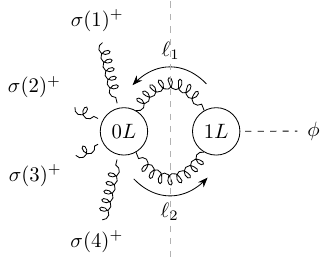}
  \hspace{0.1cm}
  \includegraphics[width=0.45\textwidth, page=4]{fig/Phi+4g-tikzext.pdf}\\
    \hspace{+0.2cm}\includegraphics[width=0.4\textwidth, page=2]{fig/Phi+4g-tikzext.pdf}
  \hspace{0.85cm}
  \includegraphics[width=0.4\textwidth, page=5]{fig/Phi+4g-tikzext.pdf}\\
      \hspace{0.21cm}\includegraphics[width=0.35\textwidth, page=3]{fig/Phi+4g-tikzext.pdf}
  \hspace{1.6cm}
  \includegraphics[width=0.35\textwidth, page=6]{fig/Phi+4g-tikzext.pdf}\\
  \caption{The double cuts of the $\phi+4g^+$ amplitude.}
  \label{fig:dcut4-sphi}
\end{figure}

Similar to lower multiplicity, the cut-constructible part of the $\phi+4g$ amplitude can be extracted by expressing the double-cut integrands in terms of contributions from double-cut scalar one-loop amplitudes. We report in~\fig{fig:dcut4-sphi} the list of the double cuts that we have to compute in the four-gluon case.

To isolate the scalar terms, we have simplified the double-cut integrands by removing spurious contributions and performing spinor-algebra manipulations. Below, we report the resulting integrands after these simplifications for all the relevant channels, with cyclic permutations implicitly understood. We divide the contributions into a first sector where we combine one-loop self-dual Higgs sub-amplitudes with tree-level Yang-Mills terms,
\allowdisplaybreaks
\begin{align}
	&\Cphiloop_{s_\phi}(\phi;1^+,2^+,3^+,4^+,\ell_1^{-},\ell_2^+)=\left[-2 A^{(0)}( \phi^\dagger;1^+,2^+,3^+,4^+)\right] \times \nonumber\\
	&\hspace{0.5cm}\frac{1}{2}\left[ \left(s_{14}s_\phi-(s_{1\phi}-s_\phi)(s_{\phi4}-s_\phi)\right) \left.I_4^{2me}(s_{\phi4},s_{1\phi};s_{23},s_\phi)\right|_{s_{\phi}\text{-cut}}\right. \nonumber \\
	&\hspace{0.5cm}\left.+(s_{\phi1}-s_{\phi})\left.I_3^{2m}(s_{1\phi},s_{\phi})\right|_{s_{\phi}\text{-cut}}+(s_{4\phi}-s_{\phi})\left.I_3^{2m}(s_{4\phi},s_{\phi})\right|_{s_{\phi}\text{-cut}}\right]\,,\\
	&\Cphiloop_{s_{1\phi}}(\phi;1^+,2^+,3^+,4^+,\ell_1^{-},\ell_2^+)=\left[-2 A^{(0)}( \phi^\dagger;1^+,2^+,3^+,4^+)\right] \times \nonumber\\
	&\hspace{0.5cm}\frac{1}{2}\left[\left(s_{14}s_\phi-(s_{1\phi}-s_\phi)(s_{4\phi}-s_\phi)\right)\left.I_4^{2me}(s_{\phi4,s_{1\phi}};s_{23},s_{\phi})\right|_{s_{1\phi}\text{-cut}}\right. \nonumber\\
	&\hspace{0.5cm}\left.+\left(s_{12}s_\phi-(s_{1\phi}-s_\phi)(s_{2\phi}-s_\phi)\right)\left.I_4^{2me}(s_{\phi2},s_{1\phi};s_{34},s_{\phi})\right|_{s_{1\phi}\text{-cut}}\right. \nonumber\\
	&\hspace{0.5cm}\left. -s_{23}s_{34} \left.I_4^{1m}(s_{23},s_{34};s_{1\phi})\right|_{s_{1\phi}\text{-cut}}+2(s_{\phi1}-s_\phi) \left.I_3^{2m}(s_{1\phi},s_\phi)\right|_{s_{1\phi}\text{-cut}}\right]\,,\\
	&\Cphiloop_{s_{12\phi}}(\phi;1^+,2^+,3^+,4^+,\ell_1^{-},\ell_2^+)=\left[-2 A^{(0)}( \phi^\dagger;1^+,2^+,3^+,4^+)\right] \times \nonumber\\
	&\hspace{0.5cm}\frac{1}{2}\left[\left(s_{12}s_\phi-(s_{1\phi}-s_\phi)(s_{4\phi}-s_\phi)\right)\left.I_4^{2me}(s_{1\phi},s_{\phi2};s_{34},s_\phi)\right|_{s_{34}\text{-cut}}\right. \nonumber\\
	&\hspace{0.5cm}\left.-s_{23}s_{34}\left.I_4^{1m}(s_{23},s_{34};s_{1\phi})\right|_{s_{34}\text{-cut}}-s_{34}s_{41}\left.I_4^{1m}(s_{34},s_{41};s_{2\phi})\right|_{s_{34}\text{-cut}}\right]\,,
\end{align}
and the remaining set of integrands,
\begin{align}
	&\CYMloop_{s_\phi}(\phi;1^+,2^+,3^+,4^+,\ell_1^{-},\ell_2^+)=-\frac{1}{6}\frac{[23]^2}{\langle 14 \rangle^2}\left[ (s_{1\phi}-s_\phi) \left. I_3^{2m}(s_\phi,s_{1\phi}) \right|_{s_{\phi}\text{-cut}} \right. \nonumber\\
	&\left.+(s_{4\phi}-s_\phi) \left. I_3^{2m}(s_\phi,s_{4\phi})\right|_{s_{\phi}\text{-cut}} +(-s_{1\phi}s_{4\phi}+s_\phi s_{23}) \left. I_4^{2me}(s_{1\phi},s_{2\phi},s_{34},s_\phi) \right|_{s_{\phi}\text{-cut}} \right] \nonumber\\
	&\left. + \frac{1}{3} \left[-2 A^{(0)}( \phi^\dagger;1^+,2^+,3^+,4^+) \right] \mathcal{C}(\phi;1^+,2^+,3^+,4^+) \left. I_2(s_\phi)\right|_{s_{\phi}\text{-cut}} \right] \label{eq:sphicut}\\
	&\CYMloop_{s_{1\phi}}(\phi;1^+,2^+,3^+,4^+,\ell_1^{-},\ell_2^+)=-\frac{1}{6}\frac{[34]^2}{\langle 12\rangle^2}\left[(s_{1\phi}-s_{34}) \left.I_3^{2m}(s_{34},s_{1\phi})\right|_{s_{1\phi}\text{-cut}} \right. \nonumber\\
	&\hspace{0.5cm}\left.+(s_{1\phi}-s_{\phi})\left.I_{3}^{2m}(s_\phi,s_{1\phi})\right|_{s_{1\phi}\text{-cut}} +(-s_{1\phi}s_{2\phi}+s_\phi s_{34})\left. I_4^{2me}(s_{1\phi},s_{2\phi};s_{34},s_\phi)\right|_{s_{1\phi}\text{-cut}} \right] \nonumber\\
	&\hspace{0.5cm}-\frac{1}{6}\frac{[23]^2}{\langle 14 \rangle^2} \left[(s_{1\phi}-s_{23}) \left.I_3^{2m}(s_{23},s_{1\phi})\right|_{s_{1\phi}\text{-cut}} +(s_{1\phi}-s{\phi})\left.I_{3}^{2m}(s_\phi,s_{1\phi}) \right|_{s_{1\phi}\text{-cut}} \right.\nonumber\\
	&\hspace{0.5cm}\left.+(-s_{1\phi}s_{4\phi}+s_\phi s_{23}) \left. I_4^{2me}(s_{1\phi},s_{4\phi};s_{23},s_\phi)\right|_{s_{1\phi}\text{-cut}} \right]\,,\\
	&\CYMloop_{s_{12\phi}}(\phi;1^+,2^+,3^+,4^+,\ell_1^{-},\ell_2^+)=-\frac{1}{6}\frac{[34]^2}{\langle 12 \rangle^2}\left[ (s_{2\phi}-s_{34})\left.I_3^{2m}(s_{34},s_{2\phi})\right|_{s_{34}\text{-cut}} \right. \nonumber\\
	&\hspace{0.5cm}\left. +(s_{1\phi}-s_{34})\left.I_3^{2m}(s_{34},s_{1\phi}) \right|_{s_{34}\text{-cut}}+(-s_{1\phi}s_{2\phi}+s_\phi s_{34})\left.I_4^{2me}(s_{1\phi},s_{2\phi};s_{34},s_\phi)\right|_{s_{34}\text{-cut}}\right]\,.
\end{align}
$I_4^{2me}(s,t;m_1^2,m_2^2)$ represents the two-mass four-point scalar function with off-shell legs, $m_1^2$ and $m_2^2$, at opposite corners of the box, while $s$ and $t$ are the usual Mandelstam variables. In~\eqn{eq:sphicut}, we have introduced an auxiliary function for the coefficient of the bubble, which reads as follows:
\begin{align}
	\mathcal{C}(\phi;1^+,2^+,3^+,4^+) = \frac{1}{2s_\phi} &\left[ \frac{-\langle 14231] +s_{12}s_{14}+s_{13}s_{14}+s_{14}^2-s_{13}s_{23}-s_{14}s_\phi}{s_{1\phi}-s_\phi}\right. \nonumber\\
	&\left.+\frac{\langle 14321]+s_{23}s_{34}-s_{14}s_\phi}{s_{4\phi}-s_\phi} \right]\,,
\end{align}
so that it simplifies and sums to one when all cyclic permutations are taken into account,
\begin{align}
	\sum_{\sigma \in S_4}\mathcal{C}(\phi;\sigma(1)^+,\sigma(2)^+,\sigma(3)^+,\sigma(4)^+) = 1\,.
\end{align}
Since the extracted discontinuities can be written in terms of those of scalar integrals, we define the cut-constructible contribution of the $\phi+4g$ amplitude as follows:
\begin{align}
	&A^{(2),cc}(\phi;1^+,2^+,3^+,4^+)= A^{(2),cc}_{\phi{(1)}\times \text{YM}{(0)}}(\phi;1^+,2^+,3^+,4^+) + A^{(2),cc}_{\phi{(0)}\times \text{YM}{(1)}}(\phi;1^+,2^+,3^+,4^+) \,,
\end{align} 
with
\begin{align}
	&A^{(2),cc}_{\phi{(1)}\times \text{YM}{(0)}}(\phi;1^+,2^+,3^+,4^+) = \left[-2 A^{(0)}( \phi^\dagger;1^+,2^+,3^+,4^+) \right] \times  \nonumber\\
	&\hspace{0.5cm}\sum_{\sigma\in \mathbb{Z}_4} \Big\{ \frac{1}{2}\left(s_{\sigma(1)\sigma(4)}s_\phi - (s_{\phi\sigma(4)}-s_\phi)(s_{\phi\sigma(1)}-s_\phi)\right) I_4^{2me}(s_{\sigma(4)\phi},s_{\phi\sigma(1)};s_{\sigma(2)\sigma(3)},s_\phi ) \nonumber\\
	&\hspace{0.5cm}- \frac{1}{2} \left( s_{\sigma(2)\sigma(3)}s_{\sigma(3)\sigma(4)} \right) I_4^{1m}(s_{\sigma(2)\sigma(3)},s_{\sigma(3)\sigma(4)};s_\phi)  + \left(s_{\phi\sigma(1)}-s_\phi\right) I_3(s_\phi,s_{\sigma(1)\phi}) \Big\} \,,\\
	&A^{(2),cc}_{\phi{(0)}\times \text{YM}{(1)}}(\phi;1^+,2^+,3^+,4^+)= \frac{1}{3}   \left[-2 A^{(0)}( \phi^\dagger;1^+,2^+,3^+,4^+) \right]  I_2(s_\phi)\nonumber\\
	&\hspace{0.5cm} -\frac{1}{6} \sum_{\sigma\in \mathbb{Z}_4} \Big\{ \frac{[\sigma(2)\sigma(3)]^2}{\langle \sigma(1)\sigma(4) \rangle^2} (-s_{\phi\sigma(1)}s_{\phi\sigma(4)}+s_\phi s_{\sigma(2)\sigma(3)}) I_4^{2me}(s_{\phi\sigma(1)},s_{\phi\sigma(4)};s_{\sigma(2)\sigma(3)},s_\phi) \nonumber \\
		&\hspace{0.5cm} + \left( \frac{[\sigma(2) \sigma(3)]^2}{\langle \sigma(1) \sigma(4) \rangle^2} +\frac{[\sigma(3)\sigma(4)]^2}{\langle \sigma(1)\sigma(2) \rangle^2}\right) (s_{\sigma(1)\phi}-s_\phi) I_3^{2m}(s_{\phi\sigma(1)}-s_\phi) \nonumber \\
	&\hspace{0.5cm} + \frac{[\sigma(2)\sigma(3)]^2}{\langle \sigma(1)\sigma(4) \rangle^2} (s_{\phi\sigma(1)}-s_{\sigma(2)\sigma(3)}) I_3^{2m}(s_{\phi\sigma(1)},s_{\sigma(2)\sigma(3)}) \nonumber\\
	&\hspace{0.5cm} + \frac{[\sigma(3)\sigma(4)]^2}{\langle \sigma(1)\sigma(2) \rangle^2} (s_{\phi\sigma(1)}-s_{\sigma(3)\sigma(4)}) I_3^{2m}(s_{\phi\sigma(1)},s_{\sigma(3)\sigma(4)}) \Big\}\,.
\end{align}
The previous decomposition in terms of scalar integrals leads to~\eqn{eq:A2ccfirstsector} and~\eqn{eq:A2ccsecondsector}.


\bibliographystyle{JHEP}
\bibliography{phiallplus_paper}

\end{document}